
\input harvmac.tex
\input epsf.tex
\noblackbox

\def\bp{\bar{\partial}}
\def\sq2{\sqrt{2}}
\def\s42{ 2^{-{1\over 4} } }

\def\p{\partial}

\def\exp{{\rm exp}}
\def\lb{\left\langle}
\def\rb{\right\rangle}
\def\ket#1{|#1\rangle}
\def\bra#1{\langle#1|}
\def\ie{i\epsilon}
\def\ra{\rightarrow}
\def\propaga#1{\left({\theta_1(#1) \over \theta_4(#1)} \right) }
\def\propp{\left({\theta'_1(0) \over \theta_4(0)}\right) }

\lref\smat{J. Polchinski, Nucl. Phys. {\bf B362} (1991) 125;
G. Moore, Nucl. Phys. {\bf B368} (1992) 557;
K. Demeterfi, A. Jevicki and J. Rodrigues, Nucl. Phys. {\bf B362} (1991)
173; G. Mandal, A. Sengupta and S. Wadia,
Mod. Phys. Lett. {\bf A6} (1991) 1465.}
\lref\discstat{V. G. Kac, in {\sl Group Theoretical
Methods in Physics}, Lecture Notes in Physics, vol. 94 
(Springer-Verlag, 1979).} 
\lref\KP{I. R. Klebanov and A. M. Polyakov,
Mod. Phys. Lett. {\bf A6} (1991) 3273.}
\lref\BPZ{A. Belavin, A. Polyakov and A. Zamolodchikov,
Nucl. Phys. {\bf B241} (1984) 333.  }
\lref\hof {D. R. Hofstadter, Phys. Rev. {\bf B14} (1976) 2239.} 
\lref\wannier{G. H. Wannier, Phys. Status Solidi  {\bf B88} (1978) 757.} 
\lref\cgcdef{C. G. Callan and D. Freed,``Phase Diagram of the Dissipative 
Hofstadter Model'', Princeton Preprint PUPT-1291, June 1991.}
\lref\dissqm{A. O. Caldeira and A. J. Leggett, Physica  {\bf 121A}(1983) 587; 
Phys. Rev. Lett. {\bf 46} (1981) 211; Ann. of Phys. {\bf 149} (1983) 374.}
\lref\osdqm{C. G. Callan, L. Thorlacius, Nucl. Phys. {\bf B329} (1990) 117.}
\lref\cardyone{J. L. Cardy,  Nuc. Phys. {\bf B240} (1984) 514.}
\lref\clny{C. G. Callan, C. Lovelace, C. R. Nappi, and S. A. Yost,
Nucl. Phys. {\bf B293} (1987) 83; Nucl. Phys. {\bf B308} (1988) 221.}
\lref\CTrivi{C. G. Callan, L. Thorlacius, Nucl.  Phys. {\bf B319} (1989) 133.}
\lref\fisher{M. P. A. Fisher and W. Zwerger, Phys. Rev. {\bf B32} (1985) 6190.}
\lref\klebanov{I. Klebanov and L. Susskind, Phys. Lett. {\bf B200} (1988) 446.}
\lref\ghm{F. Guinea, V. Hakim and A. Muramatsu,Phys. Rev. Lett.  {\bf 54} (1985) 263.}
\lref\afflud{I. Affleck and A. Ludwig, Phys. Rev. Lett {\bf 67} (1991) 161.}
\lref\tseyt{E. Fradkin and A. Tseytlin, Phys. Lett. {\bf 163B} (1985) 123.}
\lref\abou{A. Abouelsaood, C. G. Callan, C. R. Nappi and S. A. Yost,
Nucl. Phys. {\bf B280} (1987) 599.}
\lref\kondo{I. Affleck and A. Ludwig, Nucl.Phys. {\bf B360} (1991) 641.} 
\lref\caldas{C. G. Callan and S. Das,  Phys. Rev. Lett {\bf 51} (1983) 1155.}
\lref\cardytwo{J. L. Cardy, Nucl. Phys. {\bf B324} (1989) 581. }
\lref\isqm{C. G. Callan, ``Dissipative Quantum Mechanics in Particle Physics", 
Princeton preprint PUPT-1350, in ``Proceedings of the Fourth International 
Conference on Quantum Mechanics in the Light of New Technology", 
S. Kurihara ed., Japan Physical Society (1992); 
 C.L. Kane and M.P.A. Fisher, Phys Rev. Lett.{\bf 68} (1992) 1220. } 
\lref\gsw{ M. B. Green, J. H. Schwarz, and E. Witten, ``Superstring Theory'',
Cambridge University Press (1987).}
\lref\freed{D. Freed, ``Contact Terms and Duality Symmetry in the Critical 
Dissipative Hofstadter Model'', MIT preprint MIT-CTP-2170, Mar 1993,       
hep-th 9304006.} 
\lref\ishibashi{N. Ishibashi, Mod.~Phys.~Lett. {\bf A4} (1989) 251.  }
\lref\ckpap{C. G. Callan and I. R. Klebanov, Princeton University preprint
 PUPT-1432 and Institute for Advanced Studies preprint IASSNS-HEP-93/78, 
hep-th 9311092.}
\lref\ALphaseshift{I. Affleck and A. Ludwig, Phys. Rev. {\bf B48} (1993)
7297. }
\lref\affludwig{ I. Affleck and A.W.W. Ludwig, Phys. Rev. Lett. {\bf 67} (1991) 161;
   S.Eggert and I.Affleck, Phys. Rev. {\bf B 46} (1992) 10866.}
\lref\qftbook{ See any standard Quantum Field Theory book, e.g. 
C. Itzykson and J.-B. Zuber, ``Quantum Field Theory'',  McGraw-Hill (1980) }
\lref\jl{J. Polchinski and L. Thorlacius, preprint NSF-ITP-94-29,
hep-th 9404008. }

\Title{\vbox{\baselineskip12pt
\hbox{PUPT-1450}\hbox{IASSNS-HEP-94/15}\hbox{hep-th/9402113}}}
{Exact Solution of a Boundary Conformal Field Theory}
\centerline{Curtis G. Callan\footnote{$^\diamondsuit$}{callan@puhep1.princeton.edu}
\footnote{$^\clubsuit$}{On leave from Princeton University.}}
\centerline{\it School of Natural Sciences, Institute for Advanced Study}
\centerline{\it Princeton, NJ 08544}
\centerline{and}
\centerline{Igor R. Klebanov\footnote{$^\heartsuit$}
{klebanov@puhep1.princeton.edu}, Andreas W. W. Ludwig
\footnote{$^{\spadesuit}$}{ludwig@puhep1.princeton.edu} and Juan M. Maldacena
\footnote{$^\dagger$}{malda@puhep1.princeton.edu}}
\centerline{\it Department of Physics, Princeton University} 
\centerline{\it Princeton, NJ 08544} 
\vskip .3in
\centerline{\bf Abstract}
We study the conformal field theory of a free massless scalar field living
on the half line with interactions introduced via a periodic potential at the 
boundary. An $SU(2)$ current algebra underlies this system and the interacting
boundary state is given by a global $SU(2)$ rotation of the left-moving 
fields in the zero-potential (Neumann) boundary state. As the potential 
strength varies from zero to infinity, the boundary state interpolates between 
the Neumann and the Dirichlet values. The full S-matrix for scattering from
the boundary, with arbitrary particle production, is explicitly computed. 
To maintain unitarity, it is necessary to attribute a hidden 
discrete ``soliton'' degree of freedom to the boundary. The same unitarity 
puzzle occurs in the Kondo problem, and we anticipate a similar solution.
\smallskip

\Date{2/94}

\newsec{Introduction}

Conformal field theory can be defined on manifolds with boundaries, 
provided that appropriate boundary conditions are imposed \cardyone.
The Dirichlet and Neumann boundary conditions on scalar worldsheet fields
are familiar, if trivial, examples. Non-trivial examples can arise from the 
introduction of interactions localized at the boundary. In fact, many important
problems in condensed matter and particle physics, including open string 
theory \refs{\clny,\abou,\tseyt}, monopole catalysis \caldas, the Kondo model
\kondo, dissipative quantum mechanics \refs{\dissqm,\osdqm} and junctions
in quantum wires \isqm\ can be described this way. 

In this paper we will
present an exact solution of the simplest of all such theories: 
a single massless scalar field interacting through a sinusoidal potential
localized on the boundary. The period of the potential is taken to be
such that perturbatively it is a marginal operator.
If the potential strength is zero, a free 
(Neumann) boundary condition is imposed at the origin, while infinite
potential
strength leads to a  fixed (Dirichlet) condition. Both conditions
are trivially consistent with conformal invariance and the question being asked
here is whether there are conformal boundary conditions which
interpolate between them, {\it i. e.} whether our theory
is conformally invariant for arbitrary potential strength.
  Some authors have given partial evidence for
the existence of this interpolating conformal field theory \refs{\ghm,\cgcdef}, 
but a complete construction of the corresponding boundary dynamics has been 
lacking. Significant steps toward this goal were taken in a previous 
paper by two of us \ckpap, where the partition function and some 
S-matrix elements in the interpolating theory were calculated and
found to be consistent with conformal invariance.
In this paper we present a new method that 
allows us to obtain completely explicit (and very simple) expressions for 
all the dynamical quantities in the theory: partition functions, boundary 
states, S-matrix elements and so on. 

The scalar field may be compactified at any radius consistent with
the period of the potential, and one finds that all such radii are
integer multiples of the self-dual radius, $R_{sd}=\sqrt 2$.
The key to our calculation is the full exploitation of the
$SU(2)$ symmetry associated with the boundary interaction (at infinite
radius this symmetry is somewhat hidden, while at the self-dual radius
it is fully explicit).
One very interesting
feature of this $SU(2)$ is that it leads to the existence of ``soliton''
sectors (basically the $SU(2)$ partners of the usual massless boson Fock space 
states) in the scattering problem. These sectors are not visible in
naive perturbation theory and, unless they are accounted for, the S-matrix is 
non-unitary! A similar problem afflicts the
conformal field theory describing the Kondo model, and we expect
a similar resolution. 

\newsec{Exact Boundary State by SU(2) Methods}

\subsec{Setting Up The Perturbation Expansion}
Our starting point is the Lagrangian for a free field on the segment
$0<\sigma<l$, with a boundary interaction at $\sigma=0$,
\eqn\lagrangian{ L =
{1\over 8\pi}\int_0^l d\sigma (\partial_\mu X)^2-
{1\over 2}({g} e^{iX(0)/\sqrt 2}+ {\bar g} e^{-iX(0)/\sqrt 2})~. }
In order to avoid infrared complications we impose a Dirichlet boundary 
condition at $\sigma =l$ ( $X|_{\sigma=l}= 0$),
while the boundary condition at $\sigma=0$ is dynamical:
\eqn\dyn{ -{1\over \sqrt{2} \pi } {dX \over d\sigma} + i ge^{iX/\sqrt 2}
-i {\bar g} e^{-iX/\sqrt 2} =0 \ .}
The potential
strength $g$ is taken complex so that we can vary the location
of the potential minimum. By varying $|g|$ between zero and infinity, we
effectively interpolate between Neumann and Dirichlet boundary conditions 
at the origin. With the particular period chosen above, the boundary 
potential has perturbative scaling dimension one (as a boundary operator) and 
the theory is scale and conformal invariant to first order in an expansion
in the potential strength. We will show that it is {\it exactly} conformal 
invariant for any value of the potential strength and will construct
conformal boundary condition that interpolates between the limiting Neumann and
Dirichlet cases. 

Our first task is to compute the functional integral, which, as is well known 
\cardytwo, can be regarded either as an open string
partition function $Z^{BD}=tr(e^{-T H})$ or as the amplitude for free 
closed string propagation between two boundary states $Z^{BD} =
\bra B e^{- l (L_0 +\tilde{L}_0)} \ket D $ where $\ket D $ is the Dirichlet 
boundary state and $\ket B$ is the boundary state induced by the interaction. 
Expressing $\ket{B}$ as the Neumann boundary state, $\ket{N}$, acted on by the
potential term in the path integral gives the more explicit expression 
\eqn\closedstrpart{
Z^{BD} =
{\bra N} e^{-\int dt{1\over 2}( {g} e^{iX(t,0)/\sqrt 2}+
{{\bar g}} e^{-iX(t,0)/\sqrt 2} )} e^{-l (L_0 +\widetilde{L}_0 ) }
{\ket D}~.
}
In what follows we study the closed string representation, returning
to a discussion of open string physics in section 4. The expansion of 
\closedstrpart\ in powers of the potential gives
\eqn\pertexpan{\eqalign{&
Z^{BD}= Z_0^{BD}
\sum_0^{\infty}
{1 \over n!}
\lb
\prod_{i=1}^n
\int_0^T dt_i \left(
- {g\over 2} e^{iX(t_i,0)/\sqrt{2}}- {{\bar g}\over 2} e^{- iX(t_i,0)/\sqrt{2}}
\right) \rb\ , \cr
& Z_0^{BD}= {(q^2)^{-1/24} \over \sq2 f(q^2)}\sum_{n=-\infty}^\infty
(-1)^n q^{2 n^2} \ ,}
}
where $q=e^{-2\pi l/T}$ and $f(q)=\prod_{n=1}^\infty (1-q^n)$.
 The brackets are to be evaluated using the 
free-field propagator appropriate to the boundary conditions of the free problem
(periodic in time, Neumann and Dirichlet at the two spatial boundaries):
\eqn\prop{\vev{X(t_1,0) X(t_2,0)}_{\sigma=0}
=-2 \log {\theta_1^2 \bigl ({t_1-t_2\over T} | 2i\tau\bigr)
\over \theta_4^2 \bigl ({t_1-t_2\over T} | 2i\tau\bigr) }}
where $\tau=l/T$. 

In \ckpap\ it was argued that $Z^{BD}$ must have the general form 
\eqn\partclosed{Z^{BD}(g,\bar g) =
{1 \over \sq2 (q^2)^{1/24} f(q^2)}
\left (1+2\sum_{n=1}^\infty q^{n^2/2} \cos \left [{n\pi\over 2}+n\Delta
(g,\bar g)\right ]\right)~}
and explicit computation up to third order in perturbation theory gave 
$\Delta(g,\bar g)= \pi (g+\bar g)/2 + {\pi^3\over 48} (g^3 +{\bar g}^3 - 3 g^2 
{\bar g} - 3 g {\bar g}^2)+\ldots $. In this paper we rederive
eq. \partclosed\ and show how to do the 
perturbation calculation to all orders, obtaining simple analytic results in 
the process. It is convenient to rewrite \partclosed\ in the identically 
equivalent form
\eqn\partchar{\eqalign{
Z^{BD}(g,\bar g) = &\sum_{j=0, {1\over 2}, 1, \ldots} 
{\sin{(2j+1)\theta/2} \over \sin{\theta/2} } \quad
{ ((q^2)^{j^2} -(q^2)^{(j+1)^2}) \over \sq2 (q^2)^{1/24} f(q^2)} \cr
= &\sum_{j=0,{1\over 2}, 1, \ldots} 
\chi^{SU(2)}_j(\theta) {1 \over \sq2} \chi^{Vir}_{j}(q^2)~,
}}
where $\theta=2\Delta(g,\bar g) + \pi$, $ \chi^{SU(2)}_j(\theta)$ is the
spin-$j$ SU(2) character, and $\chi^{Vir}_{j}(q) \equiv f(q)^{-1} q^{-1/24}
(q^{j^2}-q^{(j+1)^2}) $ is a ``discrete state'' Virasoro character (more about 
this shortly).
What is the origin of the $SU(2)$ symmetry manifested in \partchar?
In the conformal field theory of a single free boson on the {\it open} line
(with no boundary condition), the
primary fields are: $e^{ipX}$  for any value of $p$ (weight $p^2/2$)
plus a discrete set of extra primary fields that appear at momenta 
$p=n/\sqrt{2}$ \refs{\discstat,\KP}. These discrete states are organized 
in SU(2) multiplets and can be denoted $\Phi_{j,j_z}$ where $j_z$ is related
to momentum by $j_z=p/\sqrt{2}$. They have conformal weight $h=j^2$, 
and their Virasoro characters are the $\chi^{Vir}_{j}(q)$ defined above.
It is clear from \partchar, and from the calculations in
\ckpap, that the discrete states saturate the dynamics of
our boundary problem. The reason is that the Neumann boundary state
has momentum zero, while the period of the boundary potential is such that it 
injects momenta which are integral multiples of $1/\sqrt{2}$: precisely
the values carried by the discrete states. 
Other momenta are possible, but they are not excited by the boundary 
interaction. 

The chiral SU(2) generators which transform the holomorphic discrete states 
of a given $j$ among themselves are
\eqn\sutwo{
J^{\pm}
= 
\oint
{dz\over 2 \pi i} e^{\pm i \sqrt{2} X(z)}
\;\;\;\;\;\;\;\; J^3 =\oint {dz\over 2 \pi i} {1\over \sqrt{2}} i \partial X (z)
}
(there is a corresponding set acting on the anti-holomorphic states). 
They commute with all Virasoro generators because they are contour
integrals of weight-one fields. Since $X(z)$ is the holomorphic part of the 
field $X$, $J^\pm$ are  
well-defined only when acting on states of momentum $p=n/\sqrt{2}$, which 
 are in fact the only momenta that appear in \partchar. That is why
even the $R=\infty$ theory has $SU(2)$ symmetry. 

\subsec{Exact Calculations at the Self-dual Radius}
The above remarks show that, as far as calculations of the 
partition function $Z^{BD}$ are concerned, no information 
is lost if we compactify the boson at the self-dual ($SU(2)$) 
radius, setting 
$X \sim X + 2 \pi \sqrt{2}$. The $SU(2)$ momenta $n/\sqrt 2$
are then the only ones
allowed (but we have to treat left and right momenta as independent and
we have winding states with $p_L \not= p_R$). We will now show that in this 
compactified theory, the integrals needed to evaluate the perturbation 
expansion of the partition function can be converted into global $SU(2)$ 
raising and lowering operators so that the partition function can be 
calculated by algebraic methods! Later on we will show how to
extend our simple $SU(2)$ algebraic manipulations to the infinite radius
problem.
 
Our goal is to determine the exact closed string state $\ket{B}_{SU(2)}$
corresponding to the insertion of a dynamical boundary.
As a first step, we recall that the state ${\ket N}_{SU(2)}$
which implements the Neumann boundary condition is \clny\
\eqn\neumannsutwo{
{\ket N}_{SU(2)} = \s42 e^{\sum_1^{\infty} a_r^\dagger \tilde{a}_r^\dagger}
\sum_{n=-\infty}^{\infty} {\ket{{n\over\sqrt{2}}} }_R 
{\ket{{- n\over\sqrt{2}}} }_L 
}
The normalization factor $\s42$  can be extracted by viewing the closed string 
partition function as the modular transform of the open one. It is a measure
of the zero temperature entropy of the Neumann boundary condition \affludwig.
As shown in ref. \ckpap, it is helpful to represent \neumannsutwo as a sum over
Virasoro modules: For each primary field
$\phi$ we can build a reparametrization invariant boundary state  
$\ket{\phi}\rangle=\sum_n \ket{\phi,n}\ket{\phi,\widetilde n}$
where the sum is over all the descendants of $\phi$ 
in the module \ishibashi. At the self-dual radius, the only primary
fields are the ``discrete states'' $\Phi_{j, m}\bar \Phi_{j, m'}$,
and we denote the corresponding reparametrization invariant boundary states
by $|j,m,m'\rangle\rangle$.
By examining the left and right momenta of states contributing to
\neumannsutwo, we see that only the modules with $m'=-m$ should
enter. We also note that ${\ket N}_{SU(2)}$ is a sum over all
possible  normalized (to $\s42$) oscillator states and after we change
our basis to Virasoro primaries and their descendants, every contributing
state should again be normalized to $\s42$. Therefore, we may choose
a phase convention where every contributing module enters with
coefficient  $\s42$  and rewrite the Neumann state as
\eqn\neus{
{\ket N}_{SU(2)} = \s42 \sum_{j,m}|j,m,-m\rangle\rangle ~.}
Now consider the oscillator basis version of the Dirichlet boundary state \clny:
\eqn\dirosc{
{\ket D}_{SU(2)} = \s42 e^{-\sum_1^{\infty} a_n^\dagger \tilde{a}_n^\dagger}
\sum_{n=-\infty}^{\infty} 
{\ket{{n\over\sqrt{2}}} }_R{\ket{{n\over\sqrt{2}}} }_L~. }
By the same argument as before, this can be written as a sum of  
$|j,m,m\rangle\rangle$ modules (which don't contribute to the Neumann state
except for $m=0$). Because of the minus signs in \dirosc, there will
be relative phases but, with a little thought, one can see that phase
conventions for the $|j,m,m\rangle\rangle$ can be chosen such that
\eqn\dirichletsutwo{
{\ket D}_{SU(2)}  = \s42 \sum_{j,m} e^{-i\pi j}
|j,m,m\rangle\rangle = e^{-i\pi J_1} {\ket N}_{SU(2)} ~.
}
The convention-independent fact that makes this
possible is that the $|j,0,0\rangle\rangle$ modules enter ${\ket D}_{SU(2)}$ 
and ${\ket N}_{SU(2)}$ with relative phase $(-1)^j$, a fact already 
noted in \ckpap. Remarkably, the Dirichlet state is just an $SU(2)_L$ rotation 
of the Neumann boundary state! 

At the self-dual 
radius, we can divide the field $X$ into left- and right-moving parts
\eqn\xlr{
X(t,\sigma)= x^0 + 2\pi p (t+i\sigma ) + X^{osc}
}   
\eqn\xrr{
\widetilde X(t,\sigma)= \widetilde x^0 + 2\pi \widetilde p 
(t-i\sigma ) + \widetilde 
X^{osc}
}
where $X^{osc}$ is the part that contains the oscillators,
and the left and right momenta and zero modes are independent. 
The Neumann boundary state \neumannsutwo\ reflects right-moving into
left-moving fields as follows \clny:
\eqn\reflection{
x^0 \ket{ N}=\widetilde x^0 \ket{ N}~;\quad
p \ket{ N} = -\widetilde p \ket{ N}~;\quad 
a_n \ket {N} = \widetilde{a}_{-n} \ket{N}~;\quad
\widetilde X(t,\sigma) \ket{ N}=X(t,-\sigma) \ket {N}~.
}
These properties will enable us to rewrite the action 
of the boundary perturbation in terms of $SU(2)$ generators. 

We demonstrate this by using commutativity of right- and left-movers
to rearrange the typical product of normal ordered exponentials appearing
in an $O(g^n {\bar g}^0)$ term in the expansion of 
\pertexpan\ as follows:
\eqn\reflexp{
{\bra N} \prod_{i=1}^n :e^{i (X(t_i,0)+\widetilde X(t_i,0))/\sqrt{2} }:~~=
{\bra N} \prod_{i=1}^n :e^{i \widetilde X(t_i,0)/\sqrt{2} }: 
 \prod_{i=1}^n :e^{i X(t_i,0)/\sqrt{2} }:
}
We then use \reflection\ to convert the leftmost right-moving exponential into 
a left-moving one and then commute it to the right of the remaining right-moving 
exponentials. When all the right-movers have been so eliminated, we have
 $${
{\bra N} :e^{i X(t_n,0)/\sqrt{2}}:\dots :e^{i X(t_1,0)/\sqrt{2}}:
:e^{i X(t_1,0)/\sqrt{2}}:\dots :e^{i X(t_n,0)/\sqrt{2}}:
}$$
The central pair of terms can be combined into the dimension-one current
$C:e^{i \sqrt{2} X(t_1,0)}:$ where $C$ is a normal-ordering constant that will 
be absorbed in $g$. Remember also that the the symbols $X$ and $\tilde X$
are now being used to denote the two chiral components of the field.
The integral over the time argument converts the
current to the global $SU(2)$ raising operator:
\eqn\cont{
\int_0^T dt e^{i \sqrt{2} X(t,0)} = \oint dz e^{i\sqrt{2} X(z)}
= 2\pi i J^+, \;\;\;\;\;\;\;\;\;\; z=e^{2\pi i t/T}
}
(no Jacobian is needed when changing variables from $t$ to $z$ because we are 
integrating a weight one field). We can freely move $J^+$ to the left because 
the intervening  $e^{iX/\sqrt{2}}$ operators correspond to $(j,m) = 
({1\over 2},{1\over 2})$ states and commute with the $J^+$ raising operator.
This maneuver can be repeated until all the potential insertions have
been converted to $SU(2)$ raising operators. The result for the $O(g^n\bar g^0)$ term is
\eqn\gton{
{(2 \pi i)^n \over \sq2} {1\over n!} \sum_{j',m'} \langle\langle j',m',-m'|
(J^+)^n q^{(L_0+\tilde{L}_0)} \sum_{j,m}  e^{-i\pi j}
|j,m,m\rangle\rangle \ .
}
The right-moving states are not affected by the $J^+$ operators and their 
inner product
 sets $j=j'$ and $m=m'$. The sum over descendants in the Ishibashi states 
produces a factor of $\chi_j^{Vir}(q^2) $ and all that is left is the $SU(2)$ 
matrix element ${\bra{j,{n\over 2}}}(J^+)^n {\ket{j,-{n\over 2}}}$. In the end,
we will not even need to 
calculate this matrix element, because the sum over $n$ 
generates something even simpler.

But first we have to analyze what happens to the general $g^n\bar g^m$ term in 
the expansion. In a general term, the $J^\pm$ operators will encounter
$e^{\pm i X/\sqrt{2}}$ operators with which they do {\it not} commute and we will
have to keep track of $SU(2)$ commutator terms as we reduce the amplitude.
Consider the example of an $O(g{\bar g})$ term, starting with the stage
where the first pair of $j={1\over2}$ fields have been combined into a raising
operator density:
\eqn\ggbar{\eqalign{
\int {dt_1 dt_2\over (2\pi i)^2}
{\bra N}&:e^{-i X(t_2,0)/\sqrt{2}}:: e^{i \sqrt{2} X(t_1,0)}:
:e^{-i X(t_2,0)/\sqrt{2}}: ~ =\cr
&{\bra N} J^+J^- + \int{dt_2\over 2\pi i} {\bra N}:e^{i X(t_2,0)/\sqrt{2}}::
e^{-i X(t_2,0)/\sqrt{2}}:}}
The last term comes from the commutator associated with moving 
$J_+$ past $e^{-i X /\sqrt{2}}$. The 
coincident-point product of $e^{iX/\sqrt{2}}$ with $e^{-iX/\sqrt{2}}$
evaluates to a divergent constant which can be absorbed as a constant shift in 
the interaction potential. This has no effect on the physics and can be dropped.
When we try to extend this sort of argument to higher order, we find another 
type of divergent commutator which can this time be absorbed
as a finite renormalization of the coupling strength\footnote{$^\dagger$}
{To be a bit
more specific: With a standard short distance cutoff $\epsilon$, all 
divergences can be absorbed by choosing the ``bare'' coupling constant to be 
$g/\epsilon$. The procedure we are outlining amounts to choosing the ``bare'' 
coupling to be a power series $g(1+c_1 |g|^2+\cdots)/\epsilon$. This doesn't
change the physics of the theory, but does change the precise meaning of the
parameter $g$.}. This procedure can
be generalized to all orders but we will defer the details of the argument
to Appendix A. There we will show explicitly how to 
regularize the theta function integrals so that all commutator terms 
generated by moving $J^{\pm}$ operators past $j={1\over 2}$ fields can be 
discarded.

Given such a regulator scheme, any term of order $g^n{\bar g}^m$ reduces to
the matrix element of a product of $n$ raising and $m$ lowering operators in 
some particular ordering. The sum over all terms with $n+m=N$ is easily seen
to give
\eqn\norder{
Z_n =  (\pi i)^n {1\over n!} {\bra N}_{SU(2)} 
(-g J^+ - {\bar g} J^-)^n  q^{L_0 +\widetilde{L}_0} {\ket D}_{SU(2)} 
= }$${
{(\pi i)^n \over \sq2} {1\over n!} \sum_{j,m}   {\bra{ j, m}}(-g J^+ - 
{\bar g} J^-)^n {\ket{ j, -m}} e^{-i\pi j} \chi^{Vir}_j(q^2)
}$$
The final sum over $n$ can be done explicitly, yielding
\eqn\partsum{\eqalign{
Z^{BD}= &\sum_{j=0,{1\over 2},1,\dots} \left(
\sum_{m=-j}^j {\bra{j,m}} e^{-i \pi  (g J^+ + {\bar g} J^-)} 
e^{-i\pi j}{\ket{j,-m}} \right) {1\over \sq2} \chi_j^{Vir}(q^2)\cr
=&\sum_{j=0,{1\over 2},1,\dots} \left(
\sum_{m=-j}^j {\bra{j,m}} e^{-i \pi (g J^+ + {\bar g} J^-)} 
e^{-i\pi J_1} {\ket{j,m}} \right){1 \over \sq2}  \chi_j^{Vir}(q^2)}}
where the second line is obtained using the $SU(2)$ relation 
$ e^{-i\pi J_1} {\ket{ j, m}} = e^{-i\pi j}{\ket{ j,- m}} $.

Since the sum over $m$ gives an $SU(2)$ character, it is easy to 
see that \partsum\ reduces to \partchar\ with $\Delta$ defined by
\eqn\sindelta{
\sin(\Delta(g,{\bar g}))= {(g+{\bar g}) \over 2 } {\sin(\pi |g|) 
\over |g|}~.}
This expression has a curious implication about the limit
of infinite potential strength. We expect that when $|g| \rightarrow \infty$ 
the boundary state $\ket{B}$ turns into a sum over Dirichlet states with the 
field sitting at the minima of the potential 
(whose locations are in turn set by the phase of $g$). 
This means that, in the limit of infinite potential strength, $\Delta$ 
should approach the {\it phase} of the complex coupling $g$. 
According to \sindelta, this happens at $|g|=1/2$ rather than at $|g|\to\infty$.
This is just a finite renormalization effect: As we remarked earlier, our
coupling is related to the usual one by a coupling constant redefinition of the 
form $g'= g f(g{\bar g})$\footnote{$^\dagger$}{In fact, the regularization 
used in \ckpap\ is the standard one and the coupling constant $g'$ used there 
is related to ours by $g'= g(1+{\pi^2 \over 12}g{\bar g} +\cdots)$.}.   
This can have the effect of mapping infinite coupling strength to
$|g|=1/2$ if $f$ has a singularity there.

The most important fact about \partsum\ is that, comparing with \neumannsutwo\
and \dirichletsutwo, we can read off the exact interacting boundary state:
\eqn\boundary{
\ket{B}_{SU(2)} = e^{i \pi (g J^+ + {\bar g} J^-)} 
\ket{N}_{SU(2)} = e^{i\theta^a J^a}\ket{N}_{SU(2)}~,
}
where $\vec\theta=2\pi (Re(g),-Im(g),0)$. 
The net effect of the interactions is to 
carry out a global $SU(2)$ rotation, through angle $\vec\theta$,
of the left-movers with respect to the right-movers in the original Neumann
state! Since $SU(2)$ has three generators, it may be surprising that we don't
find a three-parameter family of boundary states. Actually, there is a somewhat 
more general theory with essentially the same boundary dynamics:
We may add to \lagrangian\ a term of the form $\alpha \dot X
(\sigma=0) $ without destroying the essential features of our analysis and
the resulting $SU(2)$ rotation depends on three parameters, $Re(g)$, $Im(g)$ 
and $\alpha$. For simplicity we set $\alpha=0$ in the rest of the paper. 

$\ket{B}$ imposes reflection boundary conditions on 
$X(t, \sigma)$ which are global rotations of the conditions \reflection\
imposed by $\ket{N}$. We will discuss this in more detail in section 3.
The new state obeys, as it should, the reparametrization
invariance conditions $ (L_n - \widetilde{L}_{-n}) {\ket B} =0 $ because $\ket N$ 
obeys them and the $SU(2)$ generators commute with the Virasoro generators. 
This is a useful consistency check of our regularization procedure. 

\subsec{Extension to Other Radii and Boundary Conditions}

Now let us consider other compactification radii for the field $X$. 
The only radii consistent with the marginal potential in 
\lagrangian\ are integer multiples of the self-dual radius. We can 
extend our techniques to solve the theory at any of these radii but the
results are particularly simple in the $R=\infty$ case. Let us consider 
calculating the partition function $Z^{BD}$ at infinite radius.
At this radius the Neumann and Dirichlet boundary states have the form
\eqn\neumanninf{
{\ket N}_{R=\infty} = \s42
e^{\sum_1^{\infty} a_n^\dagger \tilde{a}_n^\dagger}
 {\ket 0}
= \s42 \sum_{j,m=0} |j,0,0\rangle\rangle \ ,
}
\eqn\dirichletinf{
{\ket D}_{R=\infty} = \s42 e^{-\sum_1^{\infty} a_n^\dagger \tilde{a}_n^\dagger}
\int dp {\ket{p} }_R
{\ket{p} }_L ~.
}
The Neumann state carries momenta $p_L=p_R=0$. 
Since the interaction term changes left and
right momentum equally and in multiples of $1/\sqrt{2}$, the only states from 
${\ket D}_{R=\infty}$ that will contribute are the ones that have the momenta
$p_L=p_R=n/\sqrt{2}$. But these are precisely the momenta that appear in the
Dirichlet state at the $SU(2)$ radius, so we can, without any error, make the 
replacement ${\ket D}_{R=\infty} \rightarrow {\ket D}_{SU(2)} $.  
We can also replace ${\ket N}_{R=\infty}$ by ${\ket N}_{SU(2)}$ in the
matrix element since the difference between the two is a collection of terms
with momenta $p_L=-p_R=n/\sqrt{2}$ with $n\ne0$, none of which have any overlap 
with the Dirichlet state (which has $p_L=p_R$). The net result is that 
$Z^{BD}_{R=\infty}=Z^{BD}_{SU(2)}$. It is easy to extend this argument to
any allowed radius and find that $Z^{BD}_R = Z^{BD}_{SU(2)}$.

The boundary state itself is a bit more interesting. The boundary state
at the $SU(2)$ radius \boundary\ has contributions from all $SU(2)$ values
of $p_L$ and $p_R$ independently. At multiples of the $SU(2)$ radius only a 
sublattice of the $SU(2)$ ~$p_L$ and $p_R$ values are allowed and one gets the 
boundary state by projecting \boundary\ onto the allowed momenta. 
At $R=\infty$, the condition is $p_L=p_R$ and we have
\eqn\projected{\eqalign{
{\ket B}_{R=\infty} = &\s42 \sum_{j} \sum_{m,n}|j,n,n \rangle\rangle
\langle\langle j,n,n| 
e^{i\theta^a J^a} |j,m,-m\rangle\rangle \cr
= & \s42 \sum_{j} \sum_{m=-j}^j {\cal D}^j_{m,-m} |j,m,m\rangle\rangle
}}
where ${\cal D}^j_{m,-m}$ is the rotation matrix element 
$$
{\cal D}^j_{m,-m} = {\bra{j,m}} e^{i \pi (g J_+ + {\bar g} J_-) } 
{\ket{j,-m}} $$
This expression is identical to the $R=\infty$ boundary state found in \ckpap. 
We do not have contributions from the continuum states because ${\ket B}$ is 
generated from a Neumann boundary state with an operator that changes the 
momentum in discrete steps and thus only the discrete momenta appear. 
For other allowed radii, ${\ket B}_R$ is obtained from ${\ket B}_{SU(2)}$ by
a similar projection: keep only the components with winding numbers
admissible at radius $R$.

Using \boundary\ we can calculate the partition function for 
other boundary conditions. For example, we can replace the Dirichlet state by 
a Neumann state. In the SU(2) radius case we get
\eqn\partnnsutwo{ Z_{SU(2)}^{BN} = 
{\bra B} q^{(L_0 + \widetilde{L}_0)} {\ket N }_{SU(2)} = \sum_{j=0,1/2,1,..}
{\sin((2j+1)\pi |g|) \over \sin(\pi |g|) }{1\over \sq2} \chi_j^{Vir}(q^2)
}
and for the infinite radius case we get 
\eqn\partnninf{
Z_{R=\infty}^{BN}={\bra B} q^{(L_0 + \widetilde{L}_0)} {\ket N } = 
\sum_{j=0,1,..} {\cal D}^j_{00}(2 \pi |g|) {1\over \sq2}
\chi_j^{Vir}(q^2)
}
where ${\cal D}^j_{00}$ is the SU(2) rotation matrix
\eqn\dzz{
{\cal D}^j_{00}(2 \pi |g|) = {1\over j!} \left( d\over {d \xi} \right)^j
\xi^j (1-\xi)^j \;\;\;\;\;\; \xi=\sin^2(\pi|g|) 
}
We have checked that these formulas agree with the first few orders 
in their respective  perturbation expansions. 

A curious result is found if both ends of the string interact
with the same sinusoidal potential at the self-dual radius.
Here the calculation of the partition function is especially simple,
$$Z_{SU(2)}^{BB}={\bra B} q^{(L_0 + \widetilde{L}_0)} {\ket B }_{SU(2)}
={\bra N}e^{-i\theta^a J^a} 
q^{(L_0 + \widetilde{L}_0)} e^{i\theta^a J^a}\ket{N}_{SU(2)}=
Z_{SU(2)}^{NN}
$$
Since the $SU(2)$ rotation commutes with the Hamiltonian and annihilates
against its adjoint, the partition
function is independent of the potential strength and the open string energy 
levels do not feel the potential at all, no matter how strong it is!
(This doesn't work for other radii where the rotation
is operated on by a projection with which it does not in general commute.)
This is reminiscent of the situation for open strings in constant electric 
fields, where attaching two equal charges to the ends of the string does not
shift the energy levels \abou. 
The same feature occurs also in the Kondo model, if one imposes two
identical Kondo boundary conditions, with a {\it non-zero
phase-shift}, at the ends of a finite strip \ALphaseshift: The resulting
spectrum does not depend on the value of the phase-shift.
Such phenomena deserve further investigation.

\newsec{Calculation and Interpretation of Scattering Amplitudes}
\subsec{A Unitarity Paradox and the Role of Solitons}
In this section we analyze the S-matrix for scattering from the dynamical 
boundary. In  \ckpap\ some exact correlation functions were derived from
the constraints of conformal invariance. We have just shown that the dynamical 
boundary state $\ket{B}$ is an $SU(2)_L$ rotation of the Neumann boundary state
$\ket{N}$. We will now show that this remarkable result, together with
the $SU(2)_1$ current algebra of the self-dual $c=1$ theory, permits an 
explicit calculation of the entire reflection S-matrix (which turns out to
be quite non-trivial and to contain some interesting lessons about solitons). 

Let us begin with a description of how one calculates the amplitude for
a collection of left-moving particles to scatter into a different collection
of right-moving particles. We have a single field $X(z,\bar z)$ 
($z=t+i\sigma$) defined on the upper half plane. Far from the boundary, which
runs along real axis, the right(left)-movers are created and destroyed by 
$\p_zX~(\p_{\bar z}X)$, and the S-matrix element is the Fourier transform 
(to pick out the desired in- and out-going energies)
of $\bra{B}\p_zX(1)\p_zX(2)\ldots\p_{\bar z}X(N)\ket{0}/\bra{B} 0\rangle$.
 If $\ket{B}$ is the 
trivial Neumann state, evaluation is simple: the boundary is eliminated 
and all the anti-holomorphic fields are converted to holomorphic fields
by the replacement $\bar\p X(t,\sigma)\to \p X(t,-\sigma)$. The resulting
purely holomorphic matrix element is computed by  treating $ X$ as a 
holomorphic free field, with propagator $\vev{X(z)X(z')}=-\log{(z-z')}$, 
defined on the whole plane. This strategy is implicit in the operator Neumann 
boundary conditions presented in \reflection. The distinction between in and 
out fields is now based on whether they are inserted above or below the real 
axis. Now turn on the boundary interactions. The principal result of the 
previous section was that the interaction terms can be rewritten as contour 
integrals of holomorphic $SU(2)$ currents along the boundary (which in fact 
sum up to a global holomorphic $SU(2)$ rotation). The anti-holomorphic fields 
$\bar\p X$ commute with these holomorphic $SU(2)$ currents and reflect through 
the underlying Neumann boundary condition into holomorphic fields in the lower 
half plane just as before. The boundary interactions can be eliminated by
closing integration contours into the lower half-plane, and carrying out the 
appropriate global $SU(2)$ rotation on every $\p X$ operator inserted in that 
half-plane. For real potential strength $g$, the explicit rotation is
\eqn\replacement{
\p X \ra \cos{(2\pi g)} \p X + \sin (2\pi g) \left[-
e^{i\sqrt{2} X} + 
e^{-i \sqrt{2} X } \over \sqrt{2}  \right]~.
}
The net result is a very simple prescription: For each incoming field, insert
a factor of $\p X$ above the real axis; for each outgoing field, insert 
the $SU(2)$-rotated version of $\p X$ below the axis; evaluate the
resulting correlator using the
free holomorphic propagator for $\p X$ and Fourier transform to pick out
specific incoming and outgoing energies. The nonlinearities introduced by
the $SU(2)$ rotation will induce arbitrarily complicated multiparticle 
scattering amplitudes. 

For the $ 1 \ra 1$ correlator this procedure is easily applied. Due to 
$X$-momentum  conservation, only the linear 
term in \replacement\ survives to give  
\eqn\onetoone{
\vev{\p X(z,{\bar z}) \bp X(z',{\bar z'})}
 =  -{\cos{2\pi g} \over (z-{\bar z'})^2}
}
Upon Fourier transforming to fix the energy we obtain 
\eqn\oto{ S(E, E') =\cos{(2\pi g)} {E\over 2} \delta (E-E')~.}
The interpretation is that the probability to scatter into only one quantum 
of the $X$-field is $\cos^2 {(2\pi g)}$ (the other factors come from our
state normalization). We can also use the above recipe to 
evaluate the $ 1 \ra n$ correlation functions 
$\vev{\bar\p X(\bar z) \p X(w_1)\ldots \p X(w_n)}$, but we find, as already
noted in  \ckpap, that they reduce to a sum of disconnected two-body pieces.
This means that the probability to scatter into more than one quantum is zero
and unitarity is violated! We note in passing that a similar
paradox arises in the study of scattering from a Kondo boundary condition
(a fact that has not, to our knowledge, been remarked upon).

The missing probability has presumably leaked into some unaccounted-for
sectors of the Hilbert space. The classical solution of this system, 
discussed in Appendix B, shows that the {\it classical}
scattering states indeed lie in 
disjoint sectors. Initially $X(t,0)$ lies at a minimum of the 
potential, but, after a wave has scattered from the boundary, $X(t,0)$ may be
displaced by an integer number of periods to a new minimum. This shift is
visible in the field $X$ as a topological soliton of winding number $n$ 
propagating away from the boundary. This suggests that 
 the scattering states should 
be labeled by an integer-valued topological charge over and above the usual 
Fock space variables indexing the number and energy of individual particles. 
The weight-one operators $\p X(z)$ we have used to describe asymptotic
particles are  only adequate to describe the trivial topological
sector. It seems reasonable that we should use the other weight-one 
operators, $e^{\pm i \sq2 X({z}) }$, to generate states in the non-zero 
charge sectors. These exponentials describe solitons (kinks) with a shift 
of $2\pi\sqrt{2}$ between the values of $X$ far to the right and far to the 
left of the soliton. Since they are holomorphic, they create excitations which 
travel in one direction at the speed of light and can have any energy (just
like the excitations created by $\p X$). Unlike the case of massive 
solitons, the profile is in no sense a unique solution of classical equations
of motion. 

The well-known arguments about whether degenerate vacuum states survive
the transition from classical to quantum physics make it less than obvious
that such soliton sectors are really present in the quantum theory.
Soliton sectors are usually associated with degenerate vacua. At the classical
level, we do have 
degeneracy associated with the multiple minima of the boundary
potential. But, the boundary quantum mechanics by itself would,
of course, have no degenerate vacua. The question is whether coupling it to
the external field theory allows 
some memory of the degenerate vacua to remain.
But the external field theory is that of a massless one-dimensional boson,
precisely the kind of theory for 
which infrared fluctuations destroy degenerate
vacua. So, the sophisticated quantum mechanic would be entitled to conjecture
that all memory of degenerate vacua on the boundary would be erased by quantum
fluctuations and that extra soliton operators would not be needed to span
the S-matrix. Our explicit construction will show that he is wrong.

With this in mind, let us look at the scattering of one incident particle into
a non-zero charge sector. Our rules tell us that the simplest amplitude for 
scattering into the charge $\pm 1$ sector is
\eqn\expon{
\vev{\bp X(\bar z) e^{\pm i \sq2 X(z') } } 
\to \pm {1\over \sq2} \sin{(2\pi g)}{1 \over (z'-{\bar z})^2}~.}
This gives a probability of $\sin^2 (2\pi g)/2$ for one charge-zero quantum 
to scatter into a singly-charged soliton state. Obviously, these two single 
soliton final states by 
themselves make up the missing probability for unitarity.
Presumably, the other possible final states, such as one soliton plus multiple 
charge zero states, 
must have zero 
probability. In order to address such questions, we need a complete 
orthonormal basis for scattering states that spans the soliton sectors. In
particular, we need to know whether a soliton-antisoliton state is a new object 
or whether it has already been included in summing over 
states built out of multiple individual charge-zero particles. We will now
recast the S-matrix calculation in such a way that the complete orthonormal
basis for the scattering states, including the soliton sectors, is clearly
identified.

\subsec{Infrared Regulators and Orthonormal Bases}
It turns out that there is an infrared subtlety that stands in the way of
identifying a complete orthonormal basis of states: processes involving solitons
are accompanied by infinitely many ordinary quanta. To do concrete S-matrix
calculations, it seems to be necessary to impose an infrared cutoff. Rather 
than putting the system in a box and losing the spatial asymptotic region 
needed to define the S-matrix, we will impose an infrared cutoff by making time 
periodic with period $T$. The cut-off restricts all energies to be multiples of 
$2 \pi \over T$ and only a finite number of states are accessible at any given 
energy. States carry discrete energy labels $n_i$ and, at the end of
any calculation, we take the continuum limit $n_i,T \ra \infty $, keeping 
$E_i = {2\pi \over T} n_i $ fixed. With this cutoff, our base space is a 
half-infinite cylinder. At $\sigma\to \infty$
 the scattering states are identified
with the discrete left- and right-moving modes of a ``closed string''
of length $T$. The dynamical boundary at $\sigma=0$ 
defines an S-matrix by the way
in which it reflects left-moving into the right-moving states. In what follows
we will use our construction of the interacting boundary state to evaluate
the S-matrix for this discretized problem.

To simplify the discussion, we consider the case where the field $X$ is 
compactified at the self-dual radius (the $S$-matrix at the infinite
radius turns out to be identical). 
The holomorphic sector has a
level-one $SU(2)$ current algebra, formed by the currents shown in \sutwo, which
can conveniently be used to classify the states. The commutation relations are 
\eqn\commutators{\eqalign{ [J_n^3, J_m^3]={n\over 2}\delta_{n+m}\ \qquad & \qquad
[J_n^+, J_m^-]= 2 J_{n+m}^3 +n \delta_{n+m}\ \cr
 [J_n^3, J_m^+]= J_{n+m}^+\ \qquad & \qquad
[J_n^3, J_m^-]= -J_{n+m}^-\ 
} }
where $J_n^\pm =J_n^1\pm J_n^2$. Since $J^3(z)=i\p X(z)/\sqrt 2$, 
the oscillator part of
the Neumann boundary state can be rewritten in terms of current operator modes:
$$ \ket{N}= \s42 \exp \sum_{n=1}^\infty {2\over n} J^3_{-n} \widetilde
J^3_{-n}\sum_{m=-\infty}^\infty\ket{m, -m}~.
$$
In Section 2, we showed that the dynamical boundary state is generated by a 
global chiral $SU(2)$ rotation acting on $\ket{N}$:
$$
\ket{B}= e^{i\theta^a J_0^a} \ket{N}~,
$$
where the rotation angle $\theta^a$ is determined by the potential strength
according to  \boundary. 
This boundary state reflects left-moving modes of the $SU(2)$
current into right-moving modes according to 
\eqn\boundarycond{
\left (\widetilde J^3_m- J^\theta_{-m}\right ) \ket{B}=0
}
where (for the case of real potential strength $g$)
\eqn\jtheta{ J^\theta_{-n}= e^{i\theta^a J_0^a} J^3_{-n}
e^{-i\theta^a J_0^a} =  \cos (2\pi g) J^3_{-n} +{\sin (2\pi g)\over 2 i}
(J^+_{-n}- J^-_{-n})
}

To calculate correlation functions, we use \boundarycond\ and \jtheta\ to 
convert a mixed product of left- and right-moving operators into a product of
operators of one chirality only. Once we have only right-moving operators, we 
replace $\ket B$ by the vacuum, $\ket 0$, the only component of $\ket B$ that 
has no left-moving excitations. Applying this recipe to the  $1\to 1$ amplitude 
in the charge zero sector gives
$$ {\bra{0} J^3_{n} \widetilde J^3_m \ket {B} \over
\bra{0} B \rangle }=
\cos (2\pi g) { \bra{0} J^3_{n} J^3_{-m} \ket {B} \over 
\bra{0} B \rangle }=
\cos (2\pi g) {n\over 2} \delta_{n-m}\ .
$$
Likewise, one finds that the $1\to n$ amplitudes vanish for $n>1$. 
Taking the continuum limit, we reproduce our previous calculation,  \oto. 
Since this approach to the calculation can be summarized by
$$\ket{out}= S \ket{in}\ , \qquad S=e^{i\theta^a J_0^a}~,$$
the S-matrix is a manifestly unitary operator. 
To verify that in detail for specific examples, we have to find
an orthonormal basis for the soliton sectors.

To describe the charge-1 sector, we may attempt to use the states of the form
\eqn\bbas{ J^+_0 J^3_{-n_1} J^3_{-n_2} \ldots J^3_{-n_i} \ket{0}~.
}
Unfortunately, because the $J^+$ operator does not commute with the
$J^3$ operators, states like this with different numbers of excitations are not 
orthogonal. The natural basis of orthogonal states is instead
\eqn\basis{
 J^3_{-n_1} J^3_{-n_2} \ldots J^3_{-n_i} \ket{q}
}
where $ \ket{q} $ is the ground state of the charge--$q$ sector. Since the
$J^3_n$ commute with each other, except for the central charge term, and since
the $\ket{q}$ are orthogonal for different $q$, this construct gives a
relativistically normalized orthogonal basis for multiparticle states in all
charge sectors. The $\ket{q}$ are well-defined  weight $q^2$ states in the 
Hilbert space of the $SU(2)$ current algebra. They can be expressed in terms of 
the action of current algebra raising operators on the charge--0 ground state:
\eqn\bas{\ket{q}= J^+_{-2q+1} J^+_{-2q+3} \ldots J^+_{-1} \ket{0}=
e^{iq\sqrt 2 X}(0) \ket{0} }
(note that the sum of the raising operator weights adds up to $q^2$ as it
should). With this construction of the basis of states, we can explicitly
calculate any S-matrix element.

\subsec{One-Body Unitarity Check and an Infrared Catastrophe}
Let us now reexamine the scattering of one  particle in the uncharged sector
 from the boundary,
using the new basis. Following the rules explained above, we find that the 
amplitude for scattering into $k$ particles in the charge-1 sector is given by
\eqn\onetomany{
\bra{ 0} J^\theta_{m}  J^3_{-n_1} \ldots J^3_{-n_k} J^+_{-1}\ket{0} =
i { \sin(2\pi g)\over 2}
 \delta_{m-1-\sum n_i}~,
}
a matrix element which depends neither on the energies nor on $k$! It is easy
to see that the amplitudes to sectors with $|q|>1$ all vanish. 

The total 
probability to scatter into a charge one final state can be written as follows
(taking into account the relativistic normalization of states \basis):
\eqn\onetochargeone{
\sum_k |\langle 1|k \rangle |^2 = {\sin^2(2\pi g)\over 2m} \sum_{k=0}^{\infty} 
{1 \over k!} \sum_{n_1\cdots n_k} {2^k \over n_1 \cdots n_k} 
\delta_{m-1-\sum n_i}~.
}
After writing the delta function as 
$$
\delta_{m-1-\sum n_i} = 
{1\over 2\pi} \int_0^{2\pi} d\theta e^{i\theta (\sum n_i +1-m)}
$$
the sums over the different $n_i$ decouple. Each sum gives a factor 
$ - \log (1- e^{i\theta} ) $ which is then raised to the power $k$ and 
summed over $k$ to give
$$
e^{-2 \log(1- e^{i\theta}) } = {1 \over (1- e^{i\theta})^2 }
$$
Next we perform the integral over $\theta$ as a contour integral surrounding 
the origin (avoiding the pole at $z=1$ by
replacing $\theta \ra \theta + i \epsilon $ in all formulae):
$$
{1\over 2\pi i}   \oint  dz{z^{-m} \over (1-z)^2 } = m~.
$$
The net result is that the total probability of going into charge one states, 
\onetochargeone, is equal to $\sin^2 (2\pi g)/ 2 $, in agreement with our 
previous argument and with unitarity. 
The advantage of this more involved calculation is that we are now
using an orthogonal basis of states, so that there is no doubt
whatsoever about proper accounting of probability.

The absence of energy dependence in the amplitudes \onetomany\
leads to a kind of ``infrared catastrophe'' for charged final states.
This tendency to produce an infinite number of soft quanta is made explicit by 
rewriting \onetomany\ in terms of the continuum variables, which gives
$$\bra{0} J^\theta_{m}  J^3_{-n_1} \ldots J^3_{-n_k} J^+_{-1}\ket{0}\sim
i { \sin(2\pi g)\over 2} {2\pi \over T} \delta(E-\sum E_i)~.
$$ 
Thus, in the limit of $T \ra \infty$ each amplitude vanishes. Indeed,
in the continuum limit it would not be consistent to have a nonzero constant 
amplitude because the probability of emitting low energy quanta would diverge
(the density of states is $\sim {1\over E}$). However, if we introduce a lower 
cut-off of order $1/T$ on the energy and sum over the probabilities, 
then the $T$-dependence disappears and we are left with a finite result (much 
like the familiar Bloch-Nordsieck calculation in QED) \qftbook. To show how it works, we 
sketch the calculation, concentrating on the $T$-dependence and neglecting 
overall factors. The total probability to scatter into $k$ particles is
\eqn\tdependence{
 \sum_k \langle 1 | k \rangle   \langle k | 1 \rangle
\sim {1\over E T^2 }
\sum_{k=0}^\infty {2^k \over k! } \int^\infty
 {dE_1 \over E_1 }\int^\infty {dE_2 \over E_2 }
\cdots \int^\infty {dE_k \over E_k } \delta(E-\sum E_i)~.
}
We decouple the integrals over final energies by using
$$
\delta(E-\sum E_i) = 
{1\over 2\pi} \int_{-\infty}^\infty dx e^{ ix(-E +\sum E_i)}
$$
and each energy integral now gives
$$
\int_{1 \over T}^\infty {d E \over E} e^{i x E} = -\log ({x\over T}) + const~.
$$
This logarithm exponentiates and cancels the factor of $1/T^2$ that multiplies
\tdependence\ . So, in the continuum limit, the probability of emitting any 
fixed number of particles is zero, but the sum over all numbers is finite. 
Thus, in a sense, the final soliton is built of an infinite number of zero 
energy particles. We should also  remark  that a physical detector will not be 
sensitive to quanta of energies lower than a certain threshold $E_{min}$, so 
when we compute any scattering cross section we will have to sum over unobserved low
energy quanta. As that sum will have a similar structure to \tdependence\ , the
$T$-dependence will disappear from the cross section (the answer will depend, 
however, on the physical threshold $E_{min}$).

\subsec{Calculations and Unitarity Checks in Higher Soliton Sectors}
The explicit check of unitarity for a single incoming quantum provides
a nice test of our new formalism.  Another non-trivial test is
the $2 \ra 2$ amplitude in the charge zero sector, 
$$ \eqalign{&{\bra{0} J^3_{n_1} J^3_{n_2} \widetilde J^3_{m_1} 
\widetilde J^3_{m_2}\ket {B} \over
\bra{0} B \rangle }=
{\cos^2 (2\pi g)\over 4} n_1 n_2 (\delta_{m_1, n_1}\delta_{m_2, n_2}+
\delta_{m_1, n_2}\delta_{m_2, n_1} )\cr
&+{\sin^2 (2\pi g)\over 4}\delta_{m_1+m_2, n_1+n_2}
 (m_1+m_2-|n_1-m_1|-|n_1-m_2|) \cr } 
$$
The first term is the disconnected product of two $1\ra 1$ scattering events
and the second is a connected intrinsic two-body scattering amplitude.
Its continuum limit
$$ {\sin^2 (2\pi g)\over 4} \delta(E_1+E_2-E'_1-E'_2)
(E_1+E_2-|E'_1-E_1|-|E'_1-E_2|)\ ,
$$
agrees with the calculation of ref. \ckpap, thus giving a further check on
the correctness of our formalism. 

We have seen that the expressions for the S-matrix elements 
depend on the basis chosen to represent the states. With a natural
orthonormal basis, the amplitudes to produce charged states
exhibit a classic infrared catastrophe associated with the proliferation
of soft quanta. Summing the probability over all the states of a given charge,
we obtain a finite physically meaningful quantity.
Below we calculate the total probability to scatter to a charge-$q$ final state
from an arbitrary multiparticle charge-zero initial state. 

In order to calculate this quantity, 
it is useful to  classify states according to their transformation
properties under the $SU(2)$ algebra generated by $J_0^a$.
For instance, any one-particle state $J^3_{-m}\ket {0}$
transforms as a $\ket {1, m}$ state. The $S$-matrix is a rotation
$e^{i\theta^a J_0^a}$ which commutes with the Casimir operator 
$J_0^aJ_0^a$. Thus, the $S$-matrix conserves the spin $J$ and rotates $J_z$ 
according to the spin $J$ representation of $e^{i\theta^a J_0^a}$. 
Hence, for a single incoming quantum the out-state is also pure spin 1, and the 
probability to have charge $q$ is given by $|{\cal D}^1_{0, q} (2\pi g)|^2$.
This formula agrees with our explicit calculations.

An $N$-particle state transforms as a product of spin-1 representations and
we need to decompose it into a sum of states with definite spin. 
Remembering that the original state is completely symmetric under particle
permutations, we find that it decomposes into a sum of states of spins 
$J=N, N-2,\cdots$. For example, a two-particle state 
$ \ket{\Psi}=J^3_{-m} J^3_{-n} \ket{0} $ can be shown to decompose as
$$\eqalign{
\ket{\Psi}~=&~\ket{\Psi_{2,0}}+\ket{\Psi_{0,0}}\cr
\ket{\Psi_{2,0}}=&\left ({2\over 3} J^3_{-m} J^3_{-n}
-{1\over 6} J^-_{-m} J^+_{-n}-{1\over 6} J^+_{-m} J^-_{-n}\right)\ket{0}\cr 
\ket{\Psi_{0,0}}=&\left ({1\over 3} J^3_{-m} J^3_{-n}
+{1\over 6} J^-_{-m} J^+_{-n}+{1\over 6} J^+_{-m} J^-_{-n}\right)\ket{0} }~.
$$
Using the current algebra, one can show that the probability to be in the 
$\ket {0, 0}$ state is
$$
{|\Psi_{0,0}|^2\over |\Psi|^2}={1\over 3}+{4\over 3 m}
$$
(we assumed that $m>n$) which reduces to $1/3$ in the continuum limit 
($m,\ n\to\infty$). The probability to be in the $\ket {2, 0}$ state
is therefore $2/3$ and the probability 
to find a final state of charge $q$ is (in the continuum limit)
$$ 
{1\over 3} \delta_{0, q}+ {2\over 3} |{\cal D}^2_{0, q} (2\pi g)|^2\ ,
$$
which is totally independent of the energies of the incoming
particles. The coefficients of the $SU(2)$ representation functions are, as 
might be expected, the squares of the Clebsch-Gordon coefficients for coupling 
two spin one representations to total spin zero and two.

For a product of $N$ $\ket {1, 0}$ representations we can write a decomposition
$$ 
J^3_{-n_1} J^3_{-n_2} \cdots J^3_{-n_N} \ket 0 =
\sum_{j=0}^N \tilde{C}^j \ket {j, 0} $$
(for even $N$, only even $j$ are present; 
for odd $N$, only odd $j$ are present).
The probability to find a charge $q$
state as a result of sending in $N$ charge  zero particles is
\eqn\genpro{ \sum_{j=0}^N |C^j|^2 |{\cal D}^j_{0, q} (2\pi g)|^2 \ ,}
where $C^j$ is the continuum limit of $\tilde{C}^j$, and $C^j$ is simply 
the Clebsch-Gordon coefficient. In the large $n$ limit, the operators 
$a_n^i= \sqrt {2\over |n|} J_n^i$ (creation and annihilation operators for 
unit-normalized particle states) have the following commutation relations
$$
[a_n^i, a_m^j] =\delta^{i, j} \delta_{m, -n}~.
$$
They have vector commutation relations with the $SU(2)$ generators as well.
Thus, the creation operators $a_n^i$ (for large $n<0$) 
act as $J=1$ c-numbers, and 
the usual decomposition of direct products of c-number tensors gives the 
Clebsch-Gordon coefficients. 

Eq. \genpro\ clearly shows that an $N$-particle charge-zero state may
scatter only into states with charges $|q|\leq N$.
An instructive special case is the scattering of
a state where $N$ particles carry equal energy $m$,
$$ 
\ket{in}= (J^3_{-m})^N\ket{0}\ ,
$$
into the maximum possible charge $q=N$. The charge-$N$ component
of the corresponding $\ket{out}$ state is 
$$\left ({\sin (2\pi g)\over 2i} J^+_{-m}\right)^N \ket{0}\ ,
$$
and we need to calculate the norm of this state and divide it
by the norm of the in-state.
The algebra is easy to do and
we find that  
$$
\tilde{C}^N = C^N \prod_{k=1}^{N-1} \left (1-{k\over m}\right )
$$ 
where $C^N$ is the Clebsh-Gordon  coefficient for coupling $N$ spin-1 states
to total spin $N$. As $n_i \ra \infty$, $\tilde C^N\ra C^N$ as expected.
The extra factors in the discretized formula ensure that enough energy is
available to create a soliton in the charge-$N$ sector: their product
vanishes for any $m$ such that $mN<N^2$, the minimum weight of
a charge-$N$ soliton.
Although a soliton of charge $N$ is
a massless particle in the continuum limit, in the discrete case its
smallest possible weight is
$N^2$. Thus, it takes a certain minimum energy to create 
such a soliton, but this energy scales to zero in the continuum limit.

\newsec{Open String Diagonalization by Adsorption Methods }

The discussion in the previous sections dealt with the boundary conformal field
theory mainly from the closed string point of view. The central object
under study was the boundary state describing the closed string state
injected into the worldsheet by the boundary interaction. It is also quite
instructive to look at this problem from an open string point of view.  We will
show that the boundary interaction can be absorbed
into the non-interacting open string Hamiltonian by quadrature, 
giving rise to a `` phase-shift'' \ALphaseshift.
The method  is a simple (though somewhat trivial)
 example of the fusion process that has been used to
solve for the multi-channel Kondo fixed point \kondo.
 We will use it
to derive expressions for the partition and correlation functions for comparison
with the closed string expressions derived in previous sections.
In order to avoid certain
technical complications, 
we study the case of real coupling $g$, with the field compactified 
at the $SU(2)$
radius and a Neumann boundary condition at the other end. We leave other
interesting cases to  Appendix C. 
 
We start with a free chiral boson $X(z)$, compactified at the $SU(2)$ 
radius and normalized so that 
$$ < X(z)X(0)> = - \ln z ~,$$
\eqn\norm{
<e^{p i X(z)}e^{-p i X(0)}> =z^{-p^2}
}
(we define $z \equiv t +i\sigma$). The momenta $p$ are restricted by the 
$SU(2)$ compactification to be integer multiples of $1/\sqrt{2}$. An
antichiral boson $\tilde X(\bar z)$ also exists and the two together describe
a free massless boson on the open line. 
We confine the system to the line 
segment $0<\sigma<l$ by imposing the Neumann condition at both ends. This
imposes the following relations between chiral and antichiral fields
$$ \tilde X(t, \sigma) = X(t, -\sigma)  
\qquad \tilde X(t, l+\sigma) = X(t, l-\sigma)~.$$
This is equivalent to a single free chiral boson $X(z)$,
satisfying periodic boundary conditions
$$\tilde X(t, \sigma) = X(t, 2l +\sigma) $$
on a circle of circumference $2l$. This is just the usual open string. 
Its partition function over a time interval $T$ is
\eqn\twisted{
 Z^{NN} = {w^{-1/24} 
\over f(w)} \sum_{Q =-\infty}^\infty w^{Q^2},
}
where $w=e^{-\pi T/l}$ and $f(w) = \prod_{n=1}^ {\infty}
(1 - w^n ) $.

Due to the periodicity of $X(z)$, the set of operators
$$ J^{3}(z) \equiv {i\over \sqrt 2} \partial_z X(z), \qquad
J^+(z) \equiv e^{ i\sqrt 2 X(z)}, \qquad
J^-(z) \equiv e^{- i\sqrt 2 X(z)}
$$
are $SU(2)_1$ Kac-Moody generators satisfying periodic boundary conditions.
We will use them to cast the Hamiltonian for this free open string theory
in Sugawara form, the most useful starting point for our subsequent
discussion of the interacting theory. Because the Neumann boundary condition is 
Virasoro invariant, the chiral and anti-chiral components of the stress tensor, 
$T = -{1\over 2}( \partial_z X)^2$ and $\tilde T =-{1\over 2}
  ( \partial_{\bar z} \tilde X)^2$, coincide 
at the boundaries. This means that the open string hamiltonian density can be 
written entirely in terms of the left-moving fields. By the usual Sugawara 
arguments for free field theories, it can be rewritten equivalently 
in terms of $U(1)$ or $SU(2)$ Kac-Moody currents:
$$ {\cal H}(\sigma) =~ :J^3(\sigma)J^3(\sigma): 
=~ :J^1(\sigma)J^1(\sigma):
= ~{ 1 \over 3 }: {\vec J}(\sigma) \cdot  {\vec J}(\sigma):~.
$$

The open string Hamiltonian can thus be written in equivalent ways 
in terms of the Fourier modes
$ J^a_n \equiv ( 1 / 2 \pi  ) \int_{-l}^{l} d\sigma 
e^{i n \pi \sigma /l} J^a(\sigma)$
of the various currents:
\eqn\hotheta{
H_0  = 
{\pi \over l} \sum_{n =-\infty}^\infty J^3_{n} J^3_{-n}=
{\pi \over l} \sum_{n =-\infty}^\infty J^1_{n} J^1_{-n}=
{\pi \over 3l} \sum_{n =-\infty}^\infty \vec J_{n}\cdot\vec J_{-n}
}
It is a straightforward matter to calculate the partition function 
corresponding to the $J^3$ Sugawara Hamiltonian: For each 
allowed value, $Q$, of the charge $J^3_0$, there is a standard 
$U(1)$ Kac-Moody module 
of conformal weight $Q^2$
 making a contribution of $w^{-1/24}f(w)^{-1}w^{Q^2}$ to
the partition function. The allowed values of $Q$ are those of the zero-mode
momenta of the string and the $SU(2)$ compactification restricts those to be 
integer. Summing over the spectrum of $Q$ gives the standard partition function
displayed in \twisted. By symmetry, the spectrum of the other $U(1)$ charge,
$J^1_0$, is the same and so is the corresponding partition function.

\subsec{ Adsorption of the Interaction: Partition Function}

Now we turn on the interaction by adding to \hotheta\
an interaction, $H_{int}$,  which has support only on the boundary $\sigma =0$.
The main point is that, since the boundary interaction is critical, the 
interaction density is a field of boundary dimension one. Since we can always
recast a conformal open string theory in terms of a single chiral field, the 
boundary interaction must be a chiral field of conformal weight one. The only
available such objects are the Sugawara currents and the straightforward 
identification is
\eqn\hintg{ 
 H_{int} =  g  J^1(\sigma =0) =
  g { \pi \over l } \sum_{n =-\infty}^\infty    J^1_n  ~.
 }
For simplicity we take $g$ to be real (in fact it turns out that the phase of 
$g$ is irrelevant for the problem at hand). If we add this to the $U(1)$ 
Sugawara Hamiltonian displayed in \hotheta, we see that the interaction can be
absorbed  by a c-number shift in the Sugawara currents:
\eqn\adsorbg{
 H_0 + H_{int} = { \pi \over l} \sum_{n =-\infty}^\infty
{\cal J}^1_{n}{\cal J}^1_{-n} + {\rm constant}
}
\eqn\redefjg{
{\cal J}^1_{n} \equiv  J^1_{n}  + g /2 ~.
}
These shift equations, evaluated at $n=0$,  show that the interaction simply 
shifts the allowed $U(1)$-charges $Q$ by $g/2$, relative to the non-interacting 
theory. The interacting Hamiltonian is the same $U(1)$ Sugawara Hamiltonian
as before, except for an overall shift of the allowed $U(1)$ charges.
The partition function is then a corresponding shift of \twisted: 
\eqn\bd{
Z^{BN} =
{w^{-1/24}\over f(w)} \sum_{Q =-\infty}^\infty w^{(Q+ { g \over 2} )^2}~.
}
When transformed to closed string variables, this reproduces
\partnnsutwo\ .

So, as far as the open string energy levels are concerned, the net effect
of the interaction 
is a rigid ``phase shift'' of the spectrum of $U(1)$ charge, which
leads to a change in the weights
of the allowed $U(1)$ Kac-Moody modules. This has
been discussed previously in the context
of the Multi-Channel Kondo model \ALphaseshift.
 It is remarkable that the
rich structure of the boundary S-matrix which we explored in the previous
sections is equivalent to such simple open string physics.

\subsec{  Adsorption of the Interaction: Boundary Condition}

The  simple adsorption method, discussed
in the last subsection, can readily be used
to obtain also the explicit boundary condition, which
the  $SU(2)$ Kac-Moody generators satisfy
at the interacting boundary. This is the open string
analogue of \boundarycond.

In order to derive this boundary condition, we
note that shifted current operators have an analytic
time evolution, $ {\cal J}^1 (t,\sigma)=$ ${\cal J}^1 (z) $,
 due to the shifted form of the Sugawara Hamiltonian
in \adsorbg.
This permits us to reintroduce a right-moving current via
$${\tilde {\cal J}}^1(t,\sigma)  = {\cal J}^1 (t,-\sigma).
$$
By definition, the boundary condition
on the $1$-component of the $SU(2)$ current is
\eqn\fourbcone{
{\cal J}^1 (t,0)=
{\tilde {\cal J}}^1(t,0) ~.
  }
Since we are only interested in the boundary condition
at the $\sigma =0$ end  of the string,
we may consider the simpler semi-infinite geometry, by
letting $l \to \infty$.

In order to find the boundary condition on the other two
components of the $SU(2)$ currents, we introduce an auxiliary
 boson field  $Y$ satisfying
\eqn\newboson{
{J}^1(z)= { i \over \sqrt{2}} \partial Y(z), \quad
{\tilde {J}}^1 (\bar z)= { i \over \sqrt{2}} {\bar \partial}
 {\tilde Y} (\bar z) ~,
} and
normalized as in \norm.
Since the  relationship between  the shifted  and
unshifted current, \redefjg,
reads in position space
\eqn\shiftposition{
{\cal J}^1(\sigma) =
{J}^1(\sigma) + \pi g \delta(\sigma) ~.
} and
since ${\cal J}^1(\sigma)$ is continuous across
the boundary, we see that the auxiliary boson field
 jumps as we cross the interacting boundary:
\eqn\newbosonshift{ 
{Y} (0^- )- Y (0^+) =   \sq2 \pi g ~.  }

Expressing the remaining components of the Kac-Moody current
in terms of  auxiliary boson field $Y$, 
$J^3 = {- 1\over 2}[ e^{i \sqrt{2} Y}+ e^{-i \sqrt{2} Y}]$, 
$J^2 = { 1\over 2i}[ e^{i \sqrt{2} Y}- e^{-i \sqrt{2} Y}]$, 
we find immediatedly the desired boundary condition:
\eqn\boundcond{
{\vec {\tilde J}}(t,\sigma) =
\pmatrix{ 1 & 0&0\cr
 	0&\cos(2 \pi g)& \sin(2 \pi g)\cr
	0&-\sin(2 \pi g)& \cos(2 \pi g)\cr}
{\vec {J}}(t,\sigma), \qquad {\rm as}\quad \sigma \to 0
}
This is the open string
 analogue of the closed string  condition in \boundarycond.


\newsec{Conclusions}
In this paper we have solved the conformal field theory of a free massless 
scalar field living on the half line and interacting via a periodic potential 
at the origin. An $SU(2)$ current algebra underlies this system and properties
of the interacting theory can be computed by $SU(2)$ methods.
As the potential strength varies from zero to infinity, the 
theory interpolates between the two trivial limits of a free field subject to
Neumann and Dirichlet boundary conditions. The intermediate conformal theories 
are quite nontrivial, having complicated multiparticle scattering amplitudes. 

Our methods allow us to compute the S-matrix in complete detail and there are
some interesting lessons to be learned from the computation. The most important
one is that scattering states are not labeled just by the number and energies
of scattered quanta: there is also a discrete charge label, reflecting the 
freedom of the field at the origin to make transitions between the different
degenerate minima of the boundary potential. These ``soliton'' sectors of the 
S-matrix are not visible in naive
perturbation theory, but must be included in order 
to maintain unitarity. Other boundary conformal theories, most 
notably those describing the overscreened Kondo model, have a similar unitarity
problem which must be resolved, we believe, by scattering into as yet
unidentified soliton sectors. A further interesting point is that there is
an infrared catastrophe associated with scattering into soliton sectors: 
infinite numbers of soft quanta carrying finite amounts of energy are always
produced.

These results should have applications in several areas. In the language of
open string theory, the boundary potential is a spacetime expectation
value of the open string tachyon field and our results should allow us to 
construct new open string theory solutions. Several condensed matter
contexts (Kondo model, quantum wires with impurities) involve effectively
one-dimensional electrons scattering from localized impurities. Our results
will describe some critical points of such systems. Finally, the original
motivation for this model came from dissipative quantum mechanics, and our
results provide a complete solution of the critical physics of the 
one-dimensional version of that problem. The simplicity of our solution
gives some hope that the more interesting critical physics of higher-dimensional
dissipative quantum mechanics models (for example the dissipative Hofstadter
model in two dimensions) can be successfully attacked. We hope to return to
these matters in future publications.
\vskip .3in
\centerline{\bf Acknowledgements}
\vskip .3in

We thank A. Yegulalp for useful discussions.
The work of C.G.C. was supported in part by DOE grant DE-FG02-90ER40542
and by the Monell Foundation. 
The work of I. R. K. was supported in part by DOE grant DE-FG02-91ER40671,
NSF Presidential Young Investigator Grant No. PHY-9157482,
James S. McDonnell Foundation Grant No. 91-48 and the A. P. Sloan Foundation. 

\vfill\eject

\bigskip
\centerline{\bf Note Added }
\bigskip

Recently, the theory solved in this paper was mapped into a fermionic
model \jl. All the fermionic calculations agree with the partition
functions we have found using the boundary state methods.
Also, \jl\ gives an explicit formula for the $R=\infty$ partition
function of open strings with identical sinusoidal potentials at the
two boundaries. This formula reveals a ``stringy'' band structure
of energy levels. Below we give a simple calculation of this effect using our 
methods.

The partition function is given in the closed string channel as
$ Z= {\bra B} q^{(L_0 + \widetilde{L}_0)} {\ket B }_{R=\infty} \ .
$
Using the formula for ${\ket B }_{R=\infty}$ (2.26), we find
$$ Z=
\sum_{j=0,1/2,1,..} {1\over \sq2} \chi_j^{Vir}(q^2)
\sum_{m=-j}^j |{\cal D}^j_{m,-m}|^2
$$
where
$$
{\cal D}^j_{m,-m} = {\bra{j,m}} e^{2i \pi g J_1 } 
{\ket{j,-m}} $$
and we have chosen $g$ to be real. As the strength of the potential varies 
from zero to infinity, $g$ varies from $0$ to $1/2$. Now,
$$\eqalign{\sum_{m=-j}^j |{\cal D}^j_{m,-m}|^2=&
\int_{-\pi}^\pi {d\phi\over 2\pi}\sum_{m=-j}^j {\bra{j,m}}
e^{i\phi J_3} e^{2i \pi g J_1 }e^{i\phi J_3}e^{-2i \pi g J_1 }{\ket{j,m}}
\cr =& \int_{-\pi}^\pi {d\phi\over 2\pi}
{\sin{(2j+1)\beta/2} \over \sin{\beta/2} }} $$
where $\beta$, the angle of the effective rotation, is given  by
$$\sin(\beta/4)=\cos(\pi g) \sin(\phi/2)
\ .$$

We may transform to the open string 
channel using the formula
$$\sum_{j=0,1/2,1,..} {1\over \sq2} \chi_j^{Vir}(q^2)
{\sin{(2j+1)\beta/2} \over \sin{\beta/2} }=
{w^{-1/24}\over f(w)} \sum_{n =-\infty}^\infty w^{(n+ { \beta \over 4\pi} )^2}
$$
and we finally find
$$Z=\int_{-\pi}^\pi {d\phi\over 2\pi}
{w^{-1/24}\over f(w)} \sum_{n =-\infty}^\infty w^{(n+ { \beta \over 4\pi} )^2}
$$
in agreement with \jl. As $\phi$ varies from $-\pi$ to $\pi$,
${\beta\over 4\pi}$ varies from $-(\half-g)$ to $(\half-g)$.
Thus, the weights of the open string Virasoro primaries are
$p^2-{1\over 24}$, and the allowed values of $p$ form symmetric bands of
width $1-2g$ centered on every integer. For $g=0$ (no potential) 
the gaps disappear and the spectrum becomes continuous; for $g=1/2$ 
(infinite potential) the spectrum  becomes discrete, as expected in the
tight binding limit.
It is remarkable that the band structure for open strings is in some
ways simpler than the band structure for particles.

				

\appendix{A}{The $SU(2)$ Regulator Scheme}

In this Appendix we exhibit the prescriptions for handling
$\theta$-function integrals which enable us to relate them to
the $SU(2)$ algebraic manipulations of section 2. As we showed there,
in performing the perturbative expansion of $Z^{BD} (g, \bar g)$
we encounter the following class of integrals,
\eqn\integral{\eqalign{
{\vev{ \epsilon_1 \ldots \epsilon_n}} \equiv 
\vev{\prod_{j=1}^n \int_0^1& d t_j e^{i\epsilon_j  X(t_j)/ \sqrt 2 } } =\cr
&\left( \theta'_1(0|2 i \tau) \over \theta_4(0|2 i \tau) \right)^n
\prod_{k=1}^n \int_0^1 d t_k \prod_{1 \leq i<j \leq n}
\left( \theta_1(t_j-t_i|2 i \tau) \over \theta_4(t_j-t_i|2 i \tau) \right)^{2
\epsilon_i \epsilon_j}
}}
where $\epsilon_i = \pm 1$. If all $\epsilon_i$ are equal, the integrals are 
finite. All other cases suffer from power law divergences, and below we 
construct a natural prescription for handling them.

{}From here on we omit the second argument of the theta functions: it will
always be $2i\tau$. The function $G(t)=({\theta_1(t)\over \theta_4(t)})^2$
plays a central role in what follows and we state some of its properties: 
It has a double zero at $t=0$ and is periodic under $t\ra t+1$ and
$t \ra t + 2i\tau $. A less obvious, but useful, symmetry property is
\eqn\inverse{G(t+i\tau) = 1/G(t)\ .}

Consider now the simplest integral of type \integral\ that needs 
regularization: $\vev{+-}$. We regularize this integral by shifting the
contour of integration away from the real axis, $t \ra t+ \ie$. Once the 
singularity at $t=0$ is avoided, we may shift the contour all the way to 
$t \ra t+ i\tau$ (we can shift it freely along the torus because of its 
periodicity properties, as long as we do not cross a pole).
\vskip 1cm
\vbox{
{\centerline{\epsfxsize=2.5in \epsfbox{x2point.eps}}}
{\centerline{\tenrm FIGURE 1: Shift of the integration contour for the $\vev{+-}$ case}}
}
\vskip .5cm
Using \inverse\ we  find
\eqn\plusminus{
\vev{+-} = \propp^2
\int_0^1 dt { \propaga{{t+i\epsilon}}}^{-2} 
 = 
\propp^2
\int_0^1 dt {\propaga{t}}^{2} 
 = \vev{++} }
which is finite. We would have obtained the same result if we had shifted the 
contour by $-\ie$, since the residue at $t=0$ is zero. 

We will regularize the general integral \integral\ by shifting the contours
of integration by an amount determined by their order in the matrix
element ($t_k \ra t_k+ \ie k$):
\eqn\npreg{
{\vev{ \epsilon_1 \cdots \epsilon_n}} \equiv
{\propp}^n
\int_0^1 d t_1 \dots \int^1_0 d t_n  \prod_{1 \leq k<l \leq n}
{\propaga{t_k-t_l +\ie (k-l)}}^{2
\epsilon_i \epsilon_j}~.
}
\vskip .7cm
\vbox{
{\centerline{\epsfxsize=1.0in \epsfbox{xnpoint.eps}}}
{\centerline{\tenrm FIGURE 2: Shift of integration contours in the 
regularized n-point function}}
}
\vskip .5cm
\noindent
In the regulated integral, shown in Fig. 2, the integrands of the individual 
contour integrals are $SU(2)$ current densities, so the $k$-th contour should
correspond to the action by the global $SU(2)$ charge $J^{\epsilon_k}$. 
If this is so, then 
the regulated integrals \npreg\ are closely related to matrix 
elements of products of the $SU(2)$ charges, the essence of the result found
in Section 2. We will show that this expectation is indeed correct.

First we note that each integral in \npreg\ is finite, and has
a well-defined limit as $\epsilon\to 0+$. It value does depend on the order
of the $t_i$'s, and in calculating the partition function it makes sense to 
average over all possible orderings. It is important to determine how the 
integral is changed by a permutation of two adjacent contours, $t_i$ and 
$t_{i+1}$. Consider therefore
\eqn\permute{\eqalign{
\vev{\epsilon_1...\epsilon_k +- \epsilon_{k+1}...\epsilon_n} = 
{\propp}^{n+2} &\int dt_1..dt_k dt dt' dt_{k+1}...dt_n \cr
G(t'-t)^{-1}\prod_{i=1}^k G(t-t_i)^{\epsilon_i}G(t-t_i)^{-\epsilon_i}
&\prod_{i=k+1}^n G(t_i-t)^{\epsilon_i}G(t_i-t)^{-\epsilon_i} F(t_1,..t_n)
}}
where $F$ contains all the propagators that do not involve the two variables
under consideration ($t$ and $t'$) and $t_k$ are complex and of the 
form $t_k = Re(t_k) + ik\epsilon$. The quantity that measures the effect of
interchanging two neighboring contours of opposite charge,
\eqn\comm{\vev{\epsilon_1...\epsilon_k +- \epsilon_{k+1}...\epsilon_n} - 
\vev{\epsilon_1...\epsilon_k -+ \epsilon_{k+1}...\epsilon_n} }
has a graphical representation as a contour integral as shown in Fig. 3.
\vskip .7cm
\vbox{
{\centerline{\epsfxsize=3.5in \epsfbox{xcommut.eps}}}
{\centerline{\tenrm FIGURE 3: Contour commutation inside an n-point function}}
}
\vskip .5cm

The difference of the two orderings reduces to a contour integral of $t'$ around 
$t$, which is given by the residue of the integrand in \permute\ at $t'=t$.
Since
\eqn\diver{
G(t'-t)^{-1} = {\left( \theta_4(0)\over \theta'_1(0) \right)}^2 {1\over (t'-t)^2}
+ {\cal O}((t'-t)^0)
}
we need to isolate the term of order $t'-t$ in the remaining factors.
After some algebra, \comm\ reduces to 
\eqn\firstinteg{\eqalign{
2\pi i {\propp}^{n} \int \prod_i dt_i dt \{&
\sum_{i=1}^k G(t-t_i)^{\epsilon_i}\partial_tG(t-t_i)^{-\epsilon_i} +\cr
&\sum_{i=k+1}^n G(t_i-t)^{\epsilon_i}\partial_tG(t_i-t)^{-\epsilon_i} \}
F(t_1,..,t_n) 
}}
In performing the integral over $t$ we may use 
\eqn\inte{
\int_0^1 dt G(t+\ie) \partial_t G(t+\ie)^{-1} = -2 \pi i
}
which finally leads to the result
$$\vev{\epsilon_1...\epsilon_k +- \epsilon_{k+1}...\epsilon_n} - 
\vev{\epsilon_1...\epsilon_k -+ \epsilon_{k+1}...\epsilon_n} =$$
$$
 -(2\pi)^2 \left( -\sum_{i=1}^k \epsilon_i + 
\sum_{i=k+1}^n \epsilon_i \right) 
{\propp}^n  \int \prod_i dt_i F(t_1,...,t_n)
$$
\eqn\commut{
= -(2\pi)^2 2 \left[\sum_{j=k+1}^n\epsilon_j -{1\over 2}
\sum_{i=1}^n \epsilon_i \right] \vev{\epsilon_1...\epsilon_n}
}
This is the formula that will enable us to relate this regularization
with the operator results of Section 2.

One may show that the quantities 
$$\vev{\epsilon_1...\epsilon_n} (2\pi i)^{-n} $$
behave under commutation of $J^{\epsilon_i}$ and
$J^{\epsilon_{i+1}}$ in the same fashion as the $SU(2)$ matrix elements
$$ {\bra{j,m}} J_{\epsilon_1}...J_{\epsilon_n} {\ket{j,{-m}}} 
\;\;\;\;\; with \;\;\;\;\;m={\sum_{i=1}^n \epsilon_i
\over 2}
$$
The contour commutation relation \commut\ 
corresponds to the $SU(2)$ relation $[J^+,J^-] = 2 J^3 $.
This is the essential reason why the perturbative evaluation
of $Z^{BD}$ can be reduced to the matrix elements of products of the
$SU(2)$ charges sandwiched between two boundary states.

Our argument can be made more precise.
First we note that there is no doubt about the validity of the
operator techniques for the case $\epsilon_i=+1$, where the integral
is convergent and equal to a matrix element of a power of $J^+$.
Now we will relate all
integrals to the integrals with $\epsilon_i=+1$.
This can be done in two steps. First we reorder the contours to bring
the integral into the form $\vev{++\ldots +--\ldots -}$. As shown above,
the extra terms picked up in the process of reordering are equivalent to
the commutator terms in the matrix element.
As a second step, in the integral $\vev{++\ldots +--\ldots -}$ we
shift the $-$ contours 
up by $i\tau$, and using  \inverse\ establish the identity
$\vev{++\ldots +--\ldots -}=\vev{++\ldots +}$.
\vskip .7cm
\vbox{
{\centerline{\epsfxsize=3.5in \epsfbox{xplusmin.eps}}}
{\centerline{Figure 4: Transformation of $\vev{+++ ... ---}$ into 
$\vev{+++...+++}$}}
}
\vskip .5cm
The same identity applies to the operator matrix elements,
$$ {\bra{j,{k-r \over 2}} } (J^+)^k (J^-)^r {\ket{j,-{k-r \over 2}}} =
{\bra{j,{k+r \over 2}} } (J^+)^{k+r} {\ket{j,-{k+r \over 2}}}
$$
Therefore, recursive reduction of the integrals
$\vev{\epsilon_1...\epsilon_n}$ to the integrals where all $\epsilon_i$
are equal, can be fully rephrased in the language of $SU(2)$
matrix elements. This is precisely what is behind the remarkable
simplicity of the results of section 2.

\appendix{B}{Classical Analysis of Scattering}

Our objective here is to gain some intuition about the role of the modes
$e^{\pm i\sq2 X}$. We note that when we scatter $\p X$ states, the constant 
background value of $X$ plays no role. Wave packets formed from $\p X$ states 
are localized perturbations that asymptote to the same constant value in 
both directions. That asymptotic constant will of course be at a minimum of 
the boundary potential. However, a localized wave packet can leave $X$ at a 
different minimum of the potential when it reflects from the boundary. The 
outgoing excitation will then be a soliton, interpolating between different
asymptotic values of the field. Since the degenerate minima of the potential
are discrete, the possible soliton charges are integer multiples of some
basic unit. The excitations in the soliton sectors are of course massless and
travel at the speed of light. This is why their quantum description is
so delicate: locally, they are indistinguishable from the usual zero-charge
excitations. 

The equation of motion derived from \lagrangian\ includes the following 
dynamical boundary condition on the field at $\sigma =0$:
\eqn\boundeqn{
{1\over \sq2} {{dX}\over d\sigma}(t,0) +2\pi  g \sin{X(t,0) \over \sq2} = 0 
}
As in the bulk $X$ satisfies the wave equation, we can express it a sum of 
left an right moving waves (with $X$ localized at a particular minimum of 
the boundary potential):  
$$X(t,\sigma)= -\sq2 \pi  + f(t+\sigma) + g(t-\sigma)~.$$ 
For an arbitrary incoming wave packet $f$, we solve for the outgoing wave 
$g$ by using \boundeqn\  and the condition $g( - \infty) = 0$. 
We see from \boundeqn\ that $X$ must evolve towards a potential minimum for 
long times, but the final minimum can be different from the initial. 

In  figure 5  we display a numerical solution of this problem for a 
particular choice of incoming wave packet such that soliton production 
actually occurs. This is an example of a final state that has to be described 
by the non-zero charge sectors of the quantum $SU(2)$ current algebra.
In the classical problem, soliton production either does or does not occur
for a given incoming wave and it is an interesting problem to identify the
incident waves for which the outgoing state switches from one value of
soliton charge to another. We have to be careful not to take the classical
calculation too literally: The period of the potential is related to the
loop expansion parameter of the quantum field theory and the particular
period for which the theory is at a conformal fixed point corresponds to 
rather strong coupling. 

{\centerline{\hbox{{\epsfxsize=2in \epsfbox{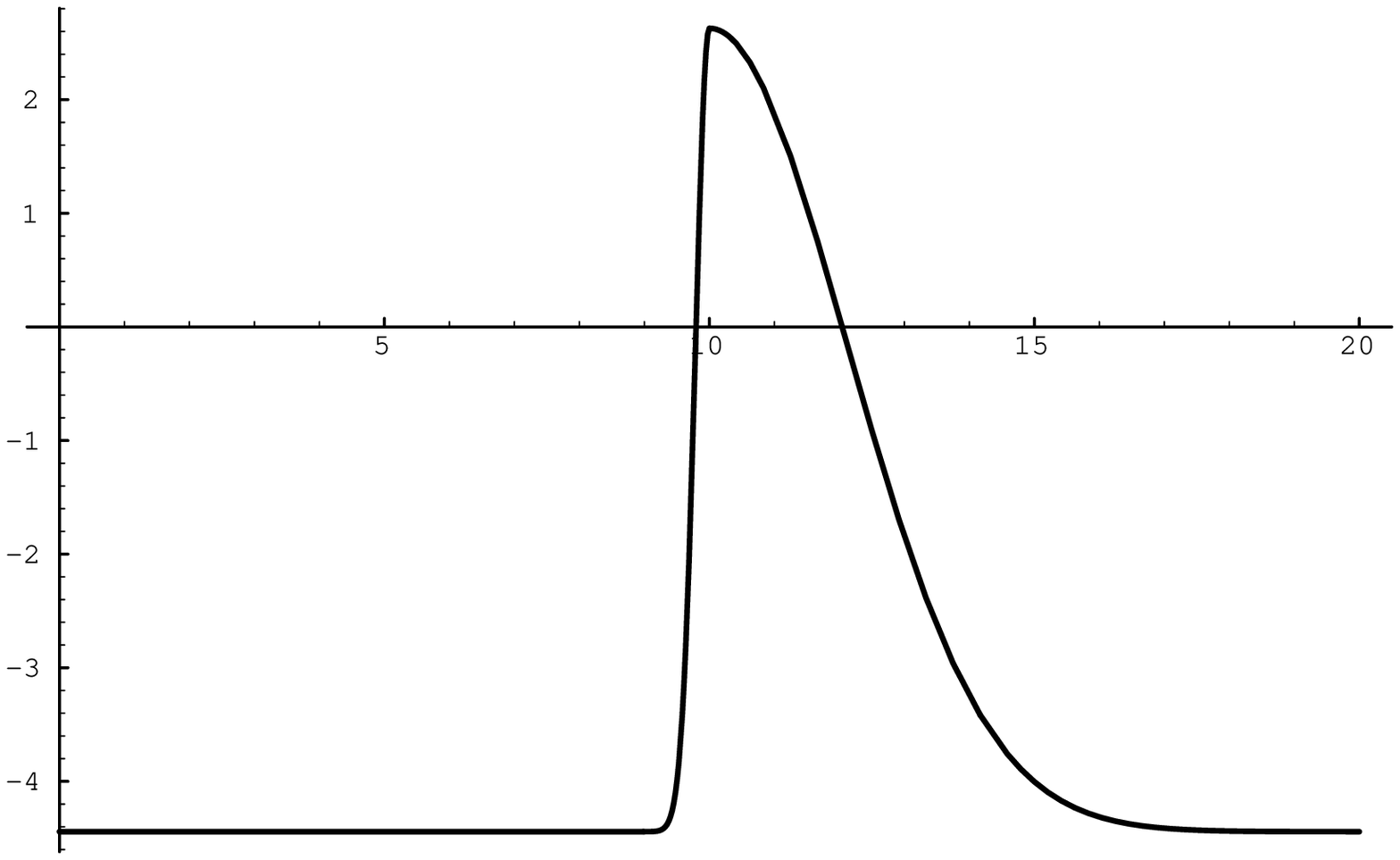}}~~~~~~~ 
{\epsfxsize=2in \epsfbox{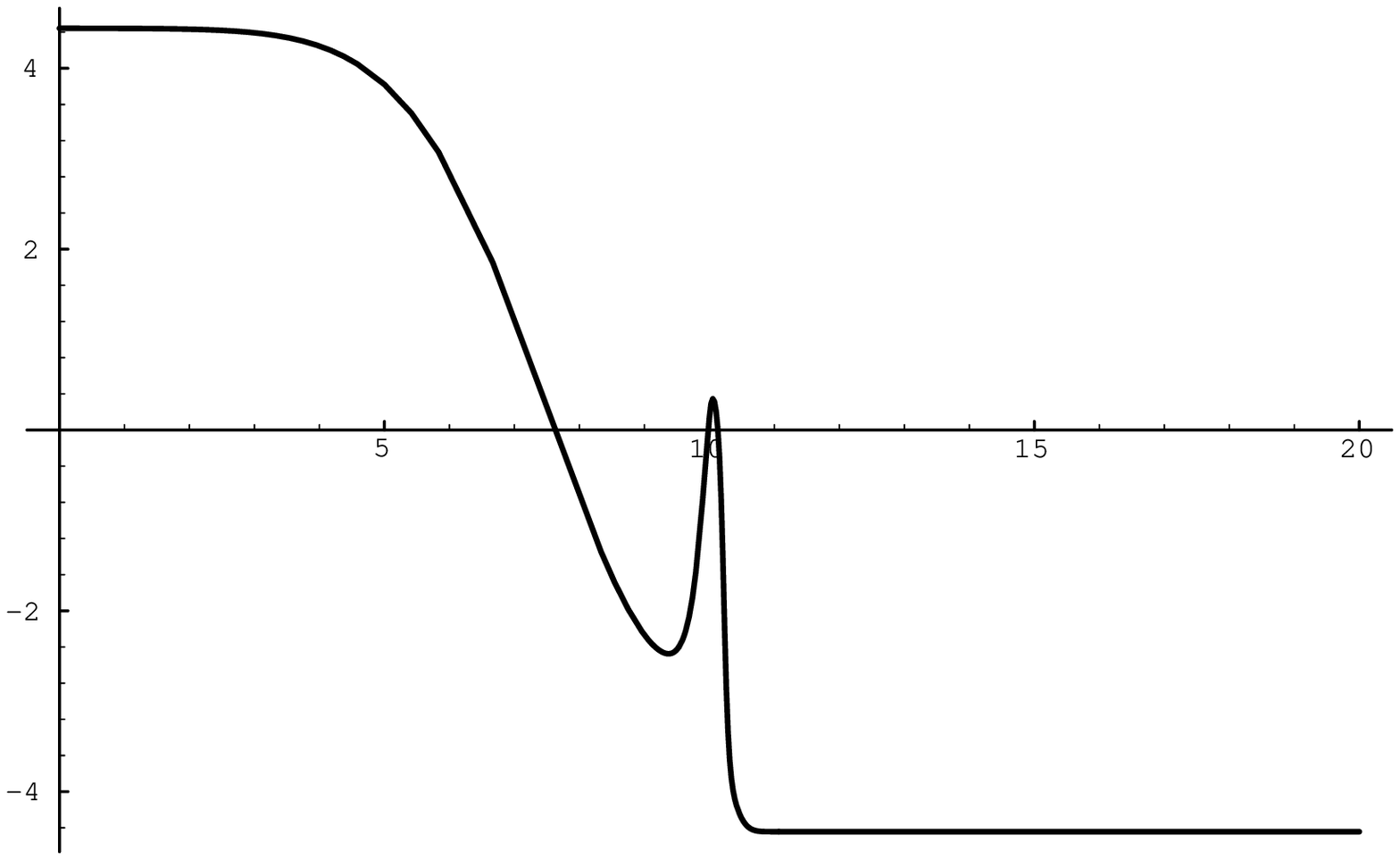}}
}}
{\centerline{\tenrm FIGURE 5: Wave packet which leaves the field at a different 
minimum after reflection.}}

\vskip .4cm

\appendix{C}{
Open String with Different Boundary Conditions at Both Ends}

In this appendix we treat the case with
interactions  added at both ends  of the open string.
In particular, we add to the  Lagrangian of the non-interacting string, 
with Neumann boundary condition on both ends, an
 interaction Lagrangian of the form
\eqn\intlagr{
L_{int}=\int_{-\infty}^{\infty} d t
\{  {\vec \theta}_1 \cdot {\vec J}(t, \sigma =0)
+
 {\vec \theta}_2 \cdot {\vec J}(t , \sigma =l ) \}~,
}
In order to investigate the effect of the presence
of the two boundary conditions, we use a conformal
mapping.
We map the strip $0 \leq {\rm Im} z \leq l$  ( recall that 
$ z= t + i\sigma $)
conformally into the upper half  complex $\zeta $ plane, 
 using
\eqn\confwz{
 \zeta  = e^{{ \pi \over l} z}
}
This maps the 
 $ \sigma = {\rm Im} z =0$ boundary
of the strip onto the  positive real $\zeta $ axis, and the
 $ \sigma = {\rm Im} z =l$ boundary
onto the  negative real $\zeta $ axis.

Due to the transformation law of the
the current operators under conformal mappings,
 ${\vec J}(z) dz =$ 
${\vec {j}}(\zeta) d\zeta  $,  
this preserves  the Neumann boundary conditions.
Furthermore,  the mapping preserves
the form of the interaction Lagrangian, 
now on the real $\zeta$ axis
in the half-infinite geometry.
However,  on the positive real axis we get
a boundary condition  with interaction parameters
${\vec \theta}_1$, and   on the negative real axis 
one with interaction parameters
${\vec \theta}_2$.

A single boundary condition of this kind has  been 
investigated in Section 4. 
We  conclude from  \boundcond~
that the currents in the half-infinite geometry
satisfy
\eqn\wbcone{
{\vec {\tilde j}}(\bar \zeta) = {\cal R}_1 
{\vec {j}}(\zeta)
}
on the positive real $\zeta$ axis, 
and 
\eqn\wbctwo{
{\vec{\tilde j}}(\bar \zeta) = {\cal R}_2^{-1} 
{\vec{j}}(\zeta)
}
on the negative real $\zeta$ axis. 
Here
${\cal R}_i$, $i=1,2$, 
  represents a rotation with angle $2 \pi  |{\vec \theta}_i|$
about the axis $ {\vec \theta}_i/|{\vec \theta}_i|$.

Both boundary conditions are conformally invariant and so
is the combination of boundary conditions  described above.
Thus we may map conformally back to the strip. This yields
\eqn\zbcone{
{\vec {\tilde J}}(\bar z) = {\cal R}_1 
{\vec {J}}(z), \qquad {\rm Im}z =0
}
\eqn\zbctwo{
{\vec {\tilde J}}(\bar z) = {\cal R}_2^{-1} 
{\vec {J}}(z), \qquad {\rm Im }z =l
}
at the two ends of the strip.

These two equations allow us again  to
eliminate the right-movers in favor of left-movers,
defined on a circle of circumference $2l$, 
which satisfy now
\eqn\striptotal{
{\vec {\tilde J}}(\sigma )=  {\cal R}_1   {\cal R}^{-1}_2
{\vec {J}}(\sigma), \qquad {\rm as} ~~ \sigma \to 0 ~.
}
This has the same form as  the boundary condition in \boundcond, 
describing a single boundary with
interaction parameter
${\vec \theta}$,  which describes the rotation
effected by  the product
$ {\cal R}_1   {\cal R}^{-1}_2 $ of
rotation matrices.
 The new parameter $\vec \theta$
is easily expressed in terms of
${\vec \theta}_1$ and
${\vec \theta}_2$ as follows
\eqn\angletotal{
\cos ( 2 \pi |{\vec \theta}|/2)
=
\cos ( 2 \pi |{\vec \theta}_1|/2) \cos ( 2 \pi  |{\vec \theta}_2|/2)
+
\sin( 2\pi  |{\vec \theta}_1|/2) \sin( 2 \pi |{\vec \theta}_2|/2)
{ {\vec \theta}_1 
\cdot 
{\vec \theta}_2 
\over  
|{\vec \theta}_1 ||{\vec \theta}_2 |}
}

We have seen in Section 4, that the partition function
under the condition 
\striptotal~
is given by \bd
\eqn\twobcpart{
Z^{B_1 B_2} =
{ 1 \over w^{1 /24} f (w) }
\sum_{n=-\infty}^{\infty }
w^{ (n+  |{\vec \theta}|/2)^2 }
}

Notice that  we may  consider the Dirichlet boundary condition $D$ 
as a special case of  $B_2$, corresponding to ${\vec{\theta}}_2 =$
$ (1/2,0,0)$.
Thus our general formula, \twobcpart~,  covers  also the
case of the partition function $Z^{BD}$, for $SU(2)$
radius, chosing ${\vec {\theta}}_1=$ $(Re(g), -Im(g),0)$.
According to the discussion following \partchar, $|{\vec \theta}|= {\Delta
\over \pi} + {1\over 2}$, with $\Delta$ given by eq. \sindelta. The open
string argument just given exactly reproduces the closed string result.

\vfill\eject
\listrefs

\bye